\documentclass[journal]{IEEEtran}
\usepackage{graphicx}
\usepackage{cite}
\usepackage{picinpar}
\usepackage{amsmath}
\usepackage{url}
\usepackage{flushend}
\usepackage[latin1]{inputenc}
\usepackage{colortbl}
\usepackage{soul}
\usepackage{multirow}
\usepackage{pifont}
\usepackage{color}
\usepackage{alltt}
\usepackage[hidelinks]{hyperref}
\usepackage{enumerate}
\usepackage{siunitx}
\usepackage{epstopdf}
\usepackage{pbox}

\usepackage[]{algorithm2e}
\usepackage{algorithmic}
\usepackage{comment}

\begin{document}
%

\title{EICO: Energy-Harvesting Long-Range Environmental Sensor Nodes with Energy-Information Dynamic Co-Optimization}

%
%
%

\author{Shitij~Avlani, Donghyun Seo, Baibhab Chatterjee,~\IEEEmembership{Student~Member,~IEEE,}
        and~Shreyas~Sen,~\IEEEmembership{Senior~Member,~IEEE}
        
\thanks{Manuscript received Month xx, 2xxx; revised Month xx, xxxx; accepted Month x, xxxx.
This work was supported in part by the SMART Films Consortium, Purdue University.}
\thanks{The authors are with the School of Electrical and Computer Engineering (ECE), Purdue University, West Lafayette, IN 47907 USA (e-mail: \{avlani, seo60, bchatte, shreyas\}@purdue.edu).}
\thanks{Color versions of one or more of the figures in this paper are available online at http://ieeexplore.ieee.org.}
\thanks{Digital Object Identifier 10.xxxx/JIOT.202x.xxxxxxx}}

\maketitle

\begin{abstract}
Intensive research on energy harvested sensor nodes with traditional battery powered devices has been driven by the challenges in achieving the stringent design goals of battery lifetime, information accuracy, transmission distance, and cost. This challenge is further amplified by the inherent power intensive nature of long-range communication when sensor networks are required to span vast areas such as agricultural fields and remote terrain. Solar power is a common energy source is wireless sensor nodes, however, it is not reliable due to fluctuations in power stemming from the changing seasons and weather conditions. This paper tackles these issues by presenting a perpetually-powered, energy-harvesting sensor node which utilizes a minimally sized solar cell and is capable of long range communication by dynamically co-optimizing energy consumption and information transfer, termed as Energy-Information Dynamic Co-Optimization (EICO). This energy-information intelligence is achieved by adaptive duty cycling of information transfer based on the total amount of energy available from the harvester and charge storage element to optimize the energy consumption of the sensor node, while employing in-sensor analytics (ISA) to minimize loss of information. This is the first reported sensor node $<35cm^2$ in dimension, which is capable of long-range communication over $>1Km$ at continuous information transfer rates of upto 1 packet/second which is enabled by EICO and ISA.
\end{abstract}

\begin{IEEEkeywords}
Energy Harvesting, energy-aware, low power, wireless sensor networks, in-sensor analytics
\end{IEEEkeywords}

\IEEEpeerreviewmaketitle

\section{Introduction}

\IEEEPARstart{A}{dvances} in semiconductor technology in the last couple of decades has enabled the proliferation of smart connected devices, collectively referred to as The Internet of Things. They have found such abundant application in all spheres of life, from smart homes and cities, wearable and implantable medical devices to agriculture and vehicles, that CISCO predicts by the year 2022 there will be machine-to-machine (M2M) communication between 14.2 billion connected devices \cite{VNI}. Low-power and cheap computing elements have enabled these devices to provide complex in-situ processing capabilities in a small and energy efficient form factor. However, a significant percentage of these devices are battery powered and require regular replacements which is bound to create a profound environmental impact, not to mention the time and cost of human intervention.

\begin{figure}[h!]
\centering
\includegraphics[width = \columnwidth]{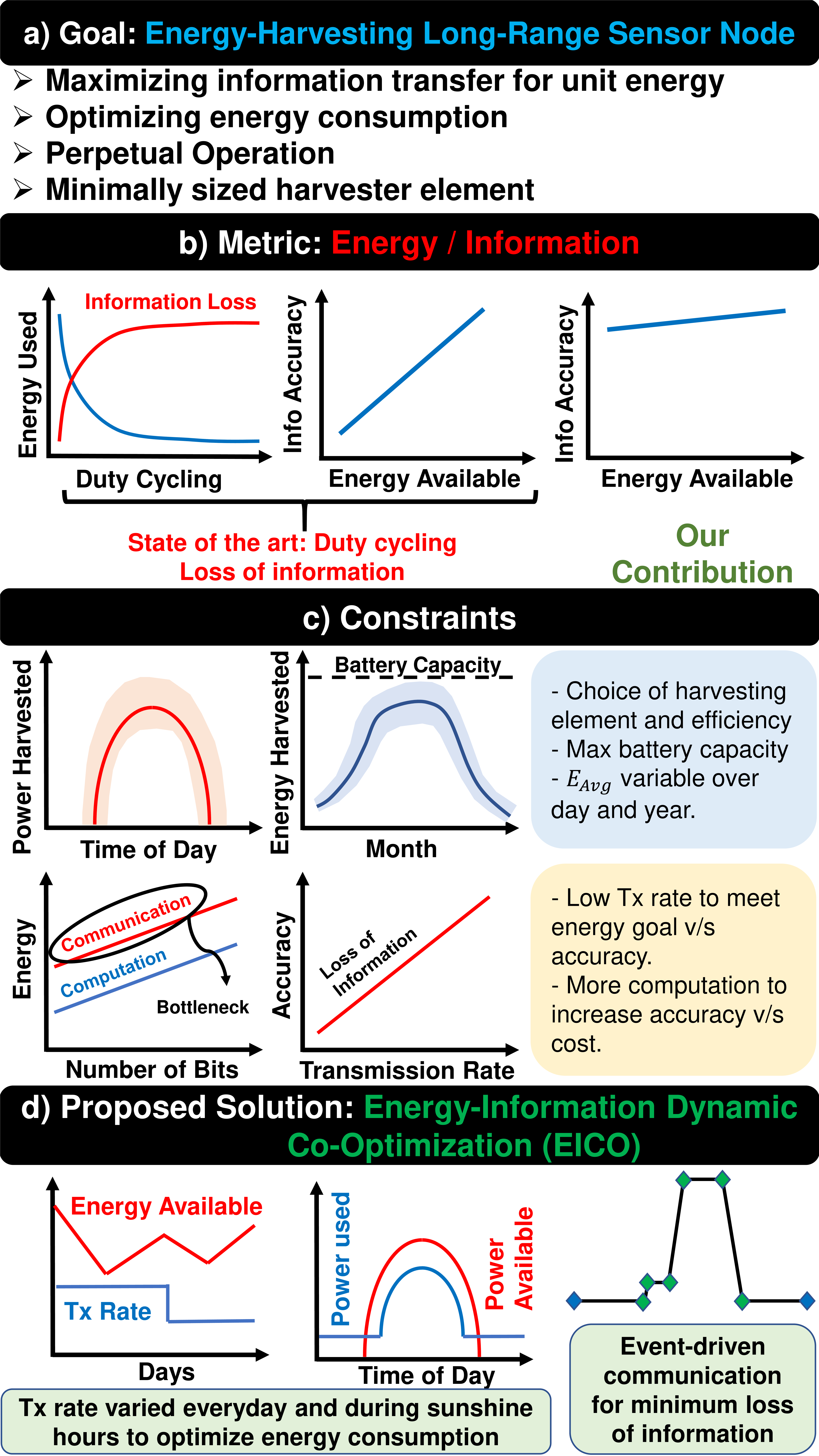}
\caption{The performance of state of the art energy-harvested long-range wireless sensor nodes is introduced in terms of an Energy/Information metric and the constraints to improve this metric are depicted. Finally, our proposed solution to optimize this metric using EICO is illustrated.}
\label{intro_figure}
\end{figure}

\subsection{Background and Motivation}
A long-range wireless sensor node is primarily used in smart cities and smart agricultural fields, and has multiple design variables ranging from the choice of transducers, power supply, communication and in-built computation capabilities, to cost, size, and network protocols. Batteries are typically the choice of power source for sensor nodes deployed in remote locations, large areas where the cost of wiring would be unfeasible, or for mobility, such as agricultural fields, habitat and environment monitoring \cite{habitat}, volcano monitoring \cite{volcano}, and structural monitoring  \cite{structural} to name a few. This creates a major limitation of finite battery capacity, resulting in a finite lifetime which adds an overhead of spending time and money to either replace the battery or place new sensor nodes while risking temporary loss of information. Designers could opt for larger batteries at the cost of increasing the size, weight and price of the device.

This has naturally sparked an increasing interest in energy harvesting sensor nodes since they can operate for many years at a time without requiring human intervention to replace the battery or the node itself. However, there are few such implementations since the instantaneous power generated by energy harvesters is not always sufficient for powering long range communication systems which consume high peak currents to support the high power output of the power amplifier to support long distances. Commercially available sensor nodes can be extremely power hungry, wherein some designs either use solar panels to meet this power requirement, which can be prohibitively expensive and large, or create low power systems that duty cycle data transmission such that information is lost in the process. Hence, its evident that there is an inconsistency between the power available from energy harvesters and the power consumed by wireless sensor nodes to perform the required task without compromising on the information reported. This motivates the creation of an energy harvesting sensor node whose goals are summarized in Fig. \ref{intro_figure}(a) capable of long range communication using a cheap and small energy source by optimizing its power consumption to function within the bounds of the energy harvester, without losing any information in the process.


\subsection{Related Work}
A multitude of software techniques have been proposed to prolong the lifetime of battery powered wireless sensor nodes without any energy harvesting modalities. Some of these methods include energy-aware network protocols, duty-cycling strategies, redundant placement of nodes, and various in-sensor analytics \cite{baibhab_jiot}, \cite{ehsn_survey}. A prominent example is an IoT device developed by Intel which implemented multiple energy-scavenging techniques like duty cycling to reduce the overall average power, however, duty cycling reduced the overall on-time \cite{ec_eg1} with an energy/information metric shown in Fig. \ref{intro_figure}(b). These methods will prolong the time between battery replacements but still require human intervention, often at the cost of information loss, sensing reliability, and increased costs due larger quantity of nodes from an increased number of hops.

Extensive research has been performed to address this problem by utilizing renewable energy through energy harvesters to power wireless sensor nodes. Some of the prominent energy sources include photovoltaics, thermoelectric generators, wind energy, piezoelectic, radio-frequency based methods, etc. \cite{wind}. Due to the low power output of these sources ($15\mu W - 30mW$ per cm under perfect conditions), most of the implementations in literature are only able to meet the needs of low-energy, short-range communication and often have low reporting intervals with loss of information as shown in Fig. \ref{intro_figure}(b). Long-range communication is extremely power intensive (150-300mW) which creates a large power discrepancy. Lee, et.al.\cite{multi-energy}, attempted to address this by proposing a floating, energy-harvested, long-range sensor node which combined solar and thermoelectric energy harvesting, but the power consumption was 6.6216 Wh/day (275.9mW) and required large solar panels to meet this demand which made the device excessively large and expensive. Stamenkovic, et.al. \cite{better_solar}, was able to shrink the size of the energy source to $40.7cm^2$ by optimizing the design using hybrid energy modelling but paid the price in information loss since data was transmitted at maximum rate of once every minute. This clearly shows that there is a discrepancy between the availability of energy from a reasonably sized energy source and the energy required to perform long range communication with minimal loss of information.  

Various methods have been explored to increase the power harvested in wireless sensor nodes by introducing a power management module to reduce the mismatch between the power harvested and the power consumed by the sensor node. These include nonlinear techniques for piezoelectric and electromagnetic energy harvesters by toggling switches at the appropriate time to form an LC oscillator using an inductor or capacitor \cite{EM} and resistive or impedance matching for maximum power transfer in energy harvesters using either a photovoltaic, thermoelectric, or piezoelectric sources \cite{piezo}, \cite{optimal_solar}. \cite{Ruan} proposed a combined power management module with an energy aware program to deal with the power mismatch by managing the energy flow from the storage capacitor. \cite{solar_prediction} proposed a solar prediction algorithm to exploit solar energy more efficiently by taking into account both the current and past-days weather conditions, however, it requires a DSP and has significant difficulties during variable weather conditions.  

To the best of our knowledge, there is no literature available on creating an energy aware system which addresses this issue by optimizing the power consumption of the sensor node by varying the transmission rate of information based on the total amount of energy available (harvested and stored), while minimizing the loss of information through event driven communication as shown in Fig. \ref{intro_figure}(d). We term this as "Energy-Information Dynamic Co-Optimization (EICO)", which has been presented in this paper leading to the first energy harvested wireless sensor node $<35cm^2$ in dimension, which is capable of long-range communication over $>1Km$ at continuous information transfer rates of up to 1 packet/second.


\subsection{Proposed Solution}

\begin{table}[h!]
\centering
\begin{tabular}{||c c||} 
\hline
Variable & Description \\ 
\hline\hline
$Tx_{minRate}$ & Minimum data transmission rate \\ 
$E_{BATT}$ & Energy currently stored in battery \\
$E_{buf}$ & Critical (buffer) energy level\\
$D_{max}$ & Lifetime (days) without energy harvested\\
$E_{Harv}$ & Energy harvested on the previous day\\
$E_{Avail}$ & Total Energy available\\
$P_{Harv}$ & Instantaneous power harvested\\
$Tx_{Rate}$ & Instantaneous data transmission rate \\ 
\hline
\end{tabular}
\vspace{1mm}
\caption{Variables for Energy-Information Dynamic Co-Optimization}
\label{energy_variable}
\end{table}

In this work, we have proposed an embedded hardware architecture and software strategies to create a perpetually powered, energy-harvested, long-range sensor node using ISA and energy-aware data transmission. ISA enables the detection of anomalies by event-driven communication and temporally compresses data to \textbf{reduce the volume of transmitted information}. Energy-aware data transmission measures the total energy available from the energy harvester on a given day and the state of the charge storage device to vary the data transmission rate of the wireless sensor node, thereby \textbf{optimizing the transfer of information to the energy consumed} by the device. This results in \textbf{increasing the transmission rate if energy available ($E_{avail}$, $P_{harv}$) is high} and reducing it in the corollary such that \textbf{perpetual operation} is maintained. A brief description of the important variables involved to enable this is shown in Table \ref{energy_variable}, with a pictorial depiction in Fig.\ref{intro_figure}. A proprietary sub-GHz transceiver from Texas Instruments \cite{TISpec} was chosen over LoRa, SigFox, and NB-IoT for long range communication since it has the best receiver sensitivity, encryption features, and provides a sufficient range (at least 1-5 Km). Additionally, it allows for the development of private networks by using the unlicensed 915MHz ISM band in Region-2 of the International Telecommunication Union.

\begin{figure}[t!]
\centering
\includegraphics[width = \columnwidth]{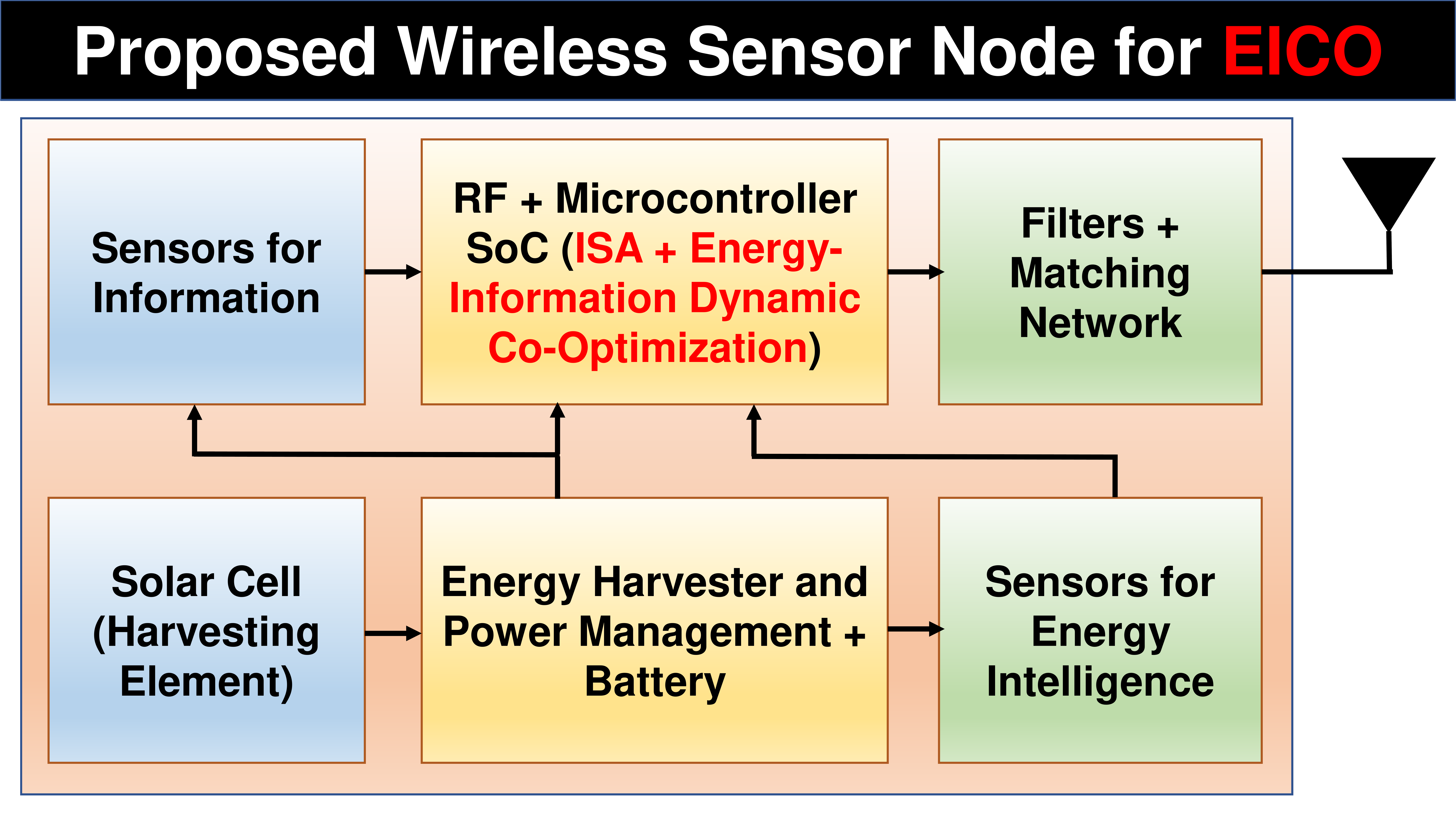}
\caption{Block diagram of the proposed energy-harvested, long-range communication wireless sensor ndoe.}
\label{block_diagram}
\end{figure}

Fig. \ref{block_diagram} shows the top-level hardware architecture of the proposed custom-built IOT sensor node. Digital sensors for temperature, humidity, and light intensity (lux) are used as the environmental sensors for information and a solar cell is used as the source for the energy harvester for demonstration purposes. The microcontroller SoC applies the ISA algorithm to the discretized and quantized values read from the sensors to detect anomalies and initiate communication when the difference between the values crosses a predefined threshold of variance. During the absence of anomalies, the sensor node duty cycles the data transmission rate which is calculated from the total energy harvested on the previous day and the energy stored in the battery, such that the device can function for at least $D_{max}$ (14 for the current implementation) days if the harvester were to fail. This will optimize the energy consumption of the device to maximize the transfer of information and improve accuracy, while ensuring that the device remains perpetually powered.

The rest of the paper flows as follows: We analyze the problem space to identify the design constraints and discuss the theoretical design trade-offs to implement an energy optimized, perpetually powered long-range sensor node. Section III describes the embedded hardware architecture of the custom-built sensor node and presents the software implementation details of the two strategies presented in this work. Section IV presents and analyzes the measurement results, and finally the paper is concluded with a summary of our contributions in Section V.


\section{Theoretical Analysis}

\subsection{Limitations of Energy Harvesting}
Advances in semiconductor technology over the years has dramatically increased the efficiency and output power of energy harvesting systems, while opening new avenues of energy sources like thermo-electric generators (TEG) and targeted radio frequency (RF) sources. TEGs can generate between \( 20\ \mu \)W to 10mW of power per $cm^{2}$ of area based on the temperature gradient. Output power from RF sources is largely limited to the \( 10\ \mu \)W - a few \( 100\ \mu \)Ws range and requires high power RF sources in close proximity. Photo-voltaic cells produce an output power ranging from \( 100\ \mu \)W - 200mW based on their construction, dimensions, spectrum of operation and light intensity. TEG and RF sources would be more suitable for devices placed in industrial locations and wearable devices, whereas photovoltaics would find better use in outdoor applications. A summary of these energy sources is shown in Table \ref{energy_source} for unit length under specific conditions.

The power intensive nature of long range communication and limited power availability warrants the need for low power architectures, careful selection of energy sources, and planning of the power budget. The sensor node proposed in this paper is primarily built for monitoring environmental variables with applications in agricultural fields or climate studies and would be placed outdoors in open fields, making solar power the obvious choice. 

\begin{table}[h!]
\centering
\begin{tabular}{||c c c||} 
\hline
Source & Power & Parameter \\ 
\hline\hline
RF & $15\mu W$ & Multiband Receiver (RF: $1mW/cm^2$)\\ 
TEG & $20.53\mu W/cm$ & Ag/Ni Thermocouple ($\Delta$T=127 C)\\
Photovoltaic & $28mW/cm^2$ & Si-Crystalline ($1KW/m^2$ solar radiation)\\
Piezoelectric & $19mW/cm^2$ & -\\
\hline
\end{tabular}
\vspace{0.5mm}
\caption{Unit power of energy sources \cite{thermo_spec} \cite{rf_spec} \cite{solar_spec}}
\label{energy_source}
\end{table}

\begin{figure}[t!]
\centering
\includegraphics[width = \columnwidth]{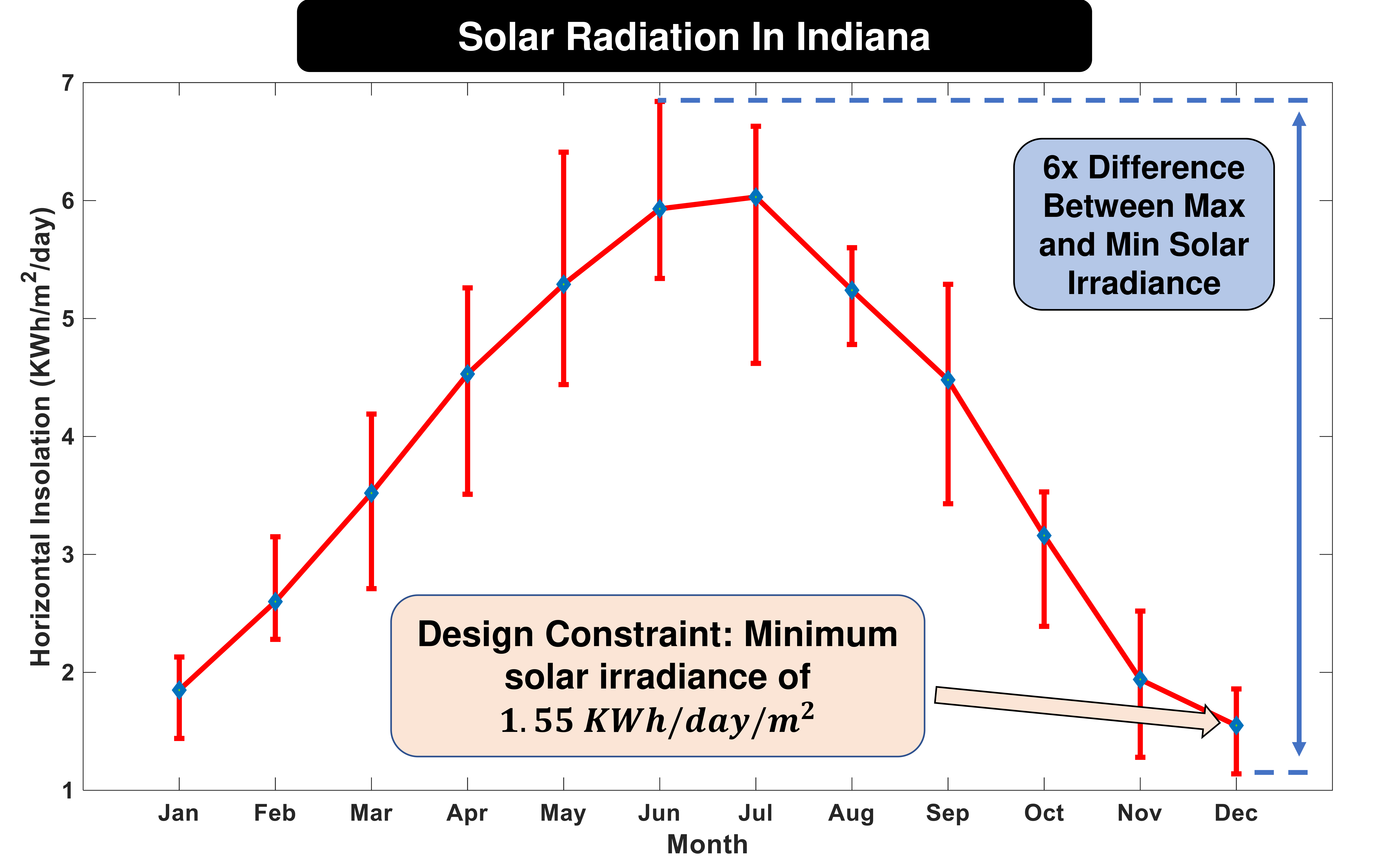}
\caption{Maximum, minimum, and average solar insolation received on an average day during a given month in Indiana which serves as the design constraint for power consumption. \cite{nasa}}
\label{solar_rad}
\end{figure}

The sensor nodes will be deployed in Indiana, which has approximately 3.5 times lower average solar insolation in winter as compared to the summer months which is illustrated in Fig. \ref{solar_rad}. The energy harvesting system and consequently the sensor node needs to be able to operate by adapting to the lower energy limit as shown in Fig. \ref{intro_figure}(c), thereby selecting that as the design constraint. As shown in Fig. \ref{solar_rad}, minimum solar insolation is in the month of December, when an average of $1.55KWh/m^{2}/day$ is received. The theoretical power harvested is then calculated as shown in Eq. (\ref{sol_pow_eqn}).

\begin{equation}
\begin{aligned}
	P_{Harvested} &= P_{rad} \times \text{Area of Solar Cell} \times \frac{\eta}{24} \\
\end{aligned}
\label{sol_pow_eqn}
\end{equation}

Where, $P_{rad}$ is the average solar power irradiated and $\eta$ is the efficiency of the cell. Assuming a dimension of $30cm^{2}$ and a conservative efficiency of 10\%($\eta$), the theoretical power consumption limit is calculated to be 2.01mW or 173.9J of energy per day. The design constraint for minimum power consumption was set at 60\% of 2.01mW, i.e. 1.20mW or 105J per day in order to to account for losses in the power management system and producing surplus energy to try and recharge the charge storage device if it is at its critical threshold.


\subsection{Theoretical Limits of Computation and Communication Energy}
Representing the energy per bit for computation and communication to be $E_{cmp,u}$ and $E_{com,u}$, respectively, the total energy consumed in a system for computation ($E_{cmp}$) and communication ($E_{com}$) is written as

\begin{equation}
\begin{aligned}
	E_{cmp} &= \left(E_{cmp,u}\right) \times \text{No. of bits switched}\\
	E_{com} &= \left(E_{com,u}\right) \times \text{No. of bits transmitted}
\end{aligned}
\label{com_eqn}
\end{equation}

\begin{figure}[b!]
\centering
\includegraphics[width=\columnwidth]{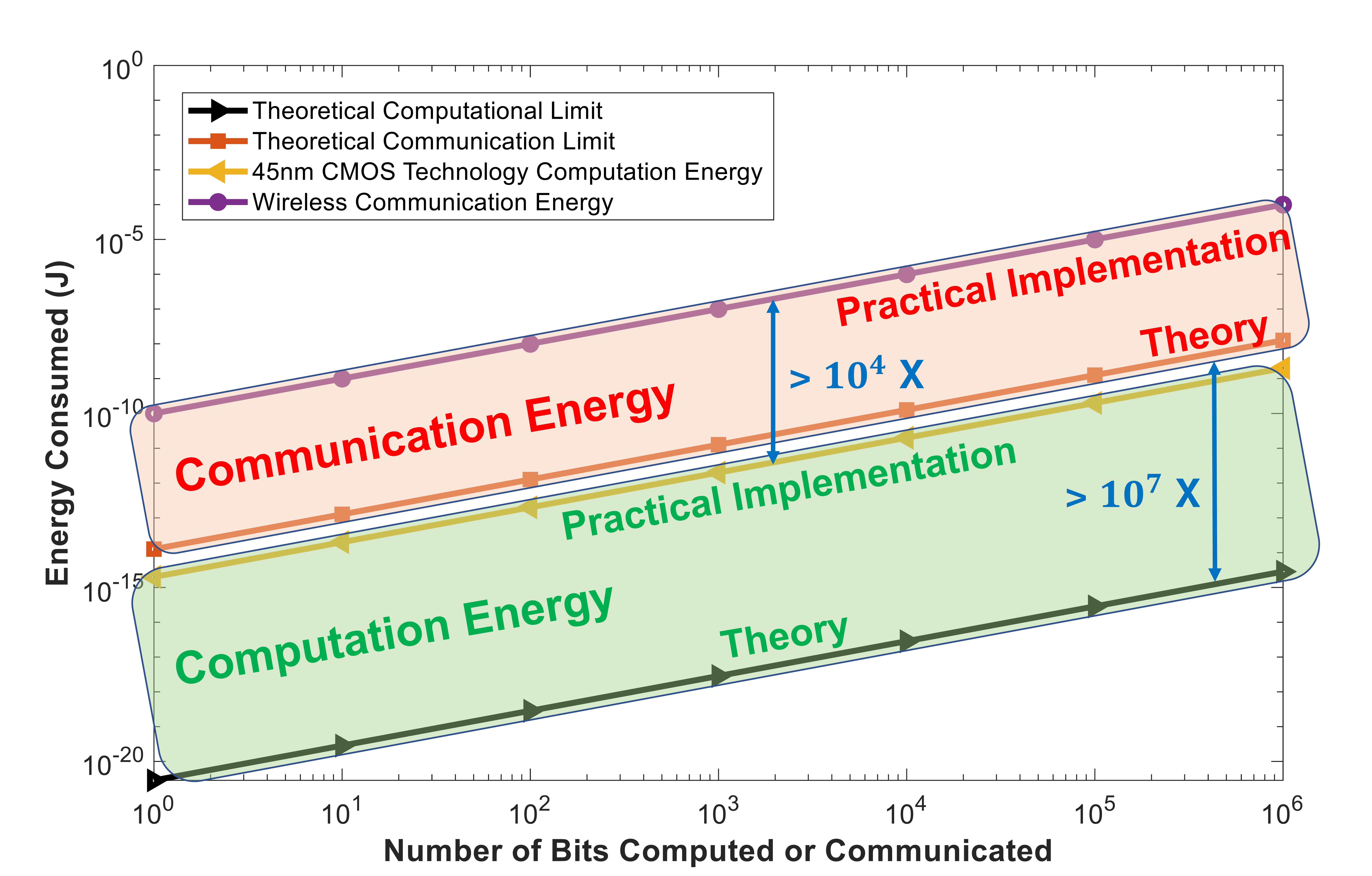}
\caption{Comparison between theoretical and practical computation and communication energies \cite{CompvsComm} \cite{30pJb} shows that computation energy is $10^4$ times less than communication energy for the same number of bits with leakage current ignored \cite{baibhab_jiot}.}
\label{comm_en}
\end{figure}

Energy consumed during computation primarily comprises of digital calculations. Therefore, it can be approximated as the dynamic energy at a frequency of operation beyond the leakage-dominant region, given by $\left(E_{cmp,u}\right) = CV^2$ \cite{NTV}. In an ideal technology that allows for zero device capacitance, $\left(E_{cmp,u}\right)$ reduces to its theoretical limit given by Landauer's principle \cite{lndr}. Eq. (\ref{lndr_eqn}) illustrates this, where $\kappa$ is Boltzmann constant and $T$ is the absolute temperature. This translates to an $(E_{cmp,u})_{th\_min}$ of $2.85 \times 10^{-21}$ J/bit at room temperatures ($T$=298K) as shown in Fig. \ref{comm_en}.

\begin{equation}
\begin{aligned}
	(E_{cmp,u})_{th\_min} &= \kappa T \times \ln{2}
\end{aligned}
\label{lndr_eqn}
\end{equation}

On the other hand, the theoretical limit of energy consumed during communication $E_{com,u}$ is given by the free-space path loss ($FSPL$) of the physical channel since the transmitter (Tx) still needs to transmit a power level which needs to be more than the receiver's (Rx) sensitivity after considering the channel loss. This is under the assumption that the receiver consumes zero power and the transmitter has a 100\% efficiency. ($FSPL$) calculated using Frii's equation \cite{Friis_eqn} \cite{Friis_link_analysis} and is shown in Eq. (\ref{friis_eqn}), where $A_{Tx}$ and $A_{Rx}$ are the antenna gains of the transmitter and receiver; $\lambda$ is the wavelength, $d$ is the distance between the transmitter and receiver, and $n$ is an empirical parameter that represents fading margin (typically between 2 to 3).

\begin{equation}
\begin{aligned}
	FSPL &= A_{Tx} . A_{Rx}(\frac{\lambda}{4{\pi}d})^{n}
\end{aligned}
\label{friis_eqn}
\end{equation}

For a typical sub-GHz protocol operating in the ISM band at $916$ MHz with $d=10$m, $FSPL$ can be estimated to be $48$ dB ($n=2,A_{Tx}=2\text{ dB, } A_{Rx}=2 \text{ dB}$). If a state-of-the-art Rx which has a sensitivity of $-120$ dBm is used in the system,then the Tx needs to transmit a minimum of $-72$ dBm. This translates to a power consumption of $63.096$ pW as theoretical minimum for power consumption. The typical data rate ($DR$) for sub-GHz communication is 5kbps. As shown in Fig. \ref{comm_en} this results in a theoretical minimum energy consumption of $(E_{com,u})_{th\_min} = 1.262 \times 10^{-14}$ J/bit, which is more than $10^7$ times higher than computational minimum given by Landauer's principle.

HDL simulations of ISA in standard 45nm CMOS process resulted in 80 $\mu$W power consumption at 100 MHz and a linear increase in computation energy at a rate of $\approx$2 fJ/bit \cite{baibhab_jiot}. The preceding discussion is summarized in Fig. \ref{comm_en} wherein the contrast between $E_{com}$ and $E_{cmp}$ is shown for the same number of bits transmitted, or switched \cite{CompvsComm}. Despite advances in wireless communication transceivers\cite{30pJb}, computation saves at least $10^4$ times more energy than communication for the same number of bits processed. This makes a strong case for incorporating ISA to process and selectively transmit data for reducing the overall system power consumption, especially when harvested energy is a scarce commodity. This conclusion is valid while the ratio between the number of bits switched during ISA and the reduction in the number of bits transmitted is less than $\left(E_{com}/E_{cmp}\right)$, which we anticipate during normal operation.


\subsection {Communication Energy and Accuracy Trade-off}
For a long-range sensor node that samples and transmits data every $N$ seconds, over $n$ seconds the communication module is on for a total time of $T_{comm} = \frac{bits \times n}{baud \times N}$. The total energy consumed during $n$ seconds is then represented by Eq. (\ref{N_eqn}).

\begin{equation}
\begin{aligned}
	E = (T_{com}.I_{com}+T_{o}.I_{cmp,lkg}+2.T_{tran}.I_{com}.\frac{n}{N})\times V
\end{aligned}
\label{N_eqn}
\end{equation}

Where $I_{com}$ is the current consumption of the communication module (along with computation of the network stack and leakage), $T_{o} = (n - T_{com})$, $I_{cmp,lkg}$ is the computation and leakage current consumed during sampling and data processing when the communication module is off, and $T_{tran}$ is the transient time during switching the module on and off (hence the factor 2) added to the initialization time. Eq. (\ref{N_eqn}) makes it evident that when $T_{com} \ll 2.T_{tran}.\frac{n}{N}$ (i.e. when $\frac{bits}{baud} \ll 2.T_{tran}$), communication energy is limited by the energy required to turn the module on or off. Conversely, when $2.T_{tran} \ll \frac{bits}{baud}$, communication energy is limited by the payload size or the number of bits transmitted.

\begin{figure}[!t]
\centering
\includegraphics[width=\columnwidth]{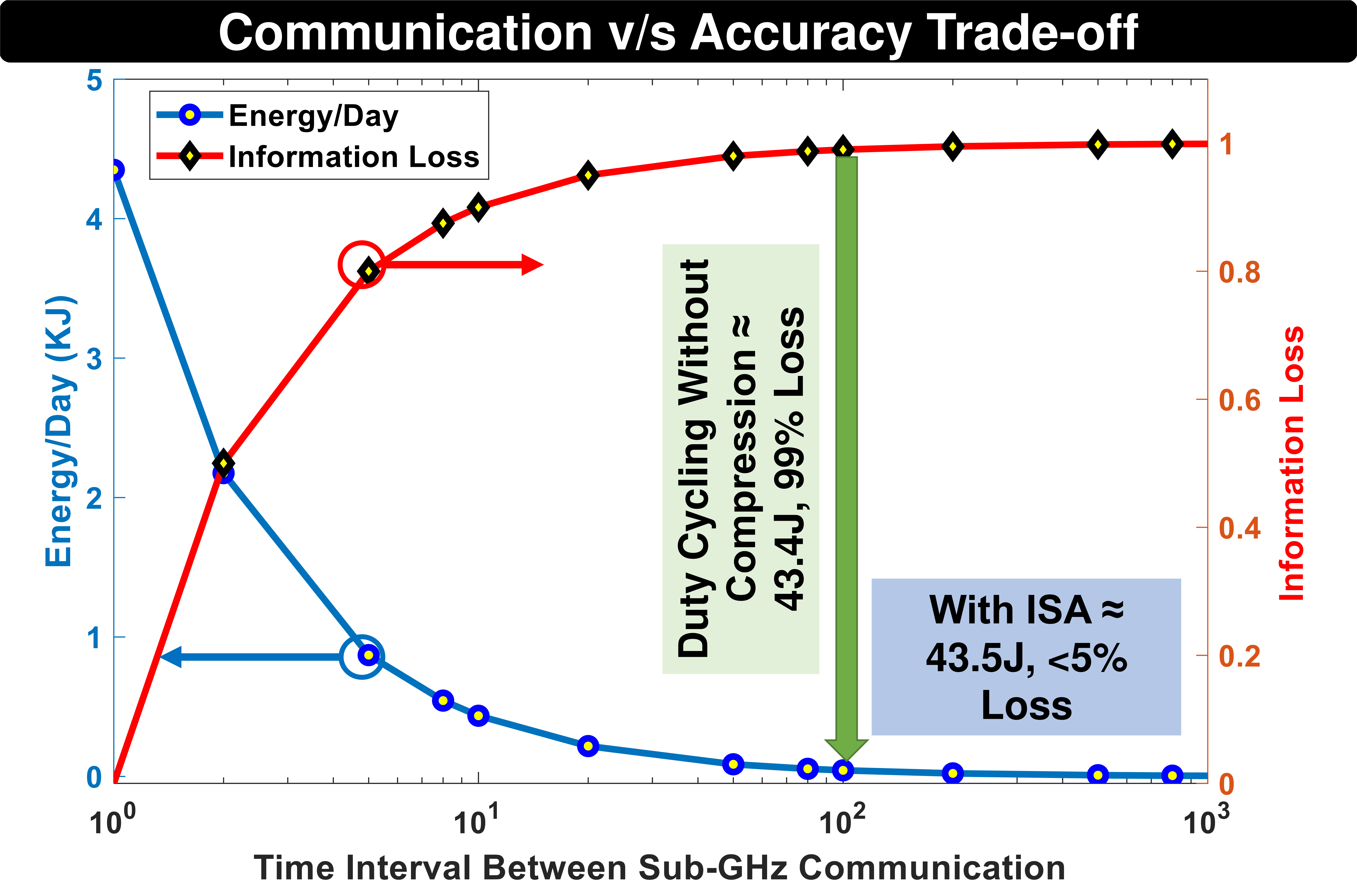}
\caption{Information loss and communication energy as a function of the time interval between sub-GHz transmission of samples, motivating the need for in-sensor analytics.}
\label{duty_N}
\end{figure}

In an effort to reduce power consumption previous methods in literature (for example, \cite{ec_eg1}, \cite{ec_eg2}) preferred a duty cycle based approach to limit the amount of switching energy by increasing $N$.This results in an increased probability of losing important, useful information. Fig. \ref{duty_N} illustrates this scenario by graphically depicting the communication energy per day and rate of information lost as a function of $N$. The values were measured for sub-GHz communication by comparing $N=100$ to a baseline of $N=1$, wherein we see a $50$X reduction in energy consumed at the cost of 99\% loss of information. Implementing a mechanism to avoid these losses while taking an acceptable hit in energy consumption makes a strong case for utilizing finely tuned in-sensor analytics and energy aware adaptive transmission.  


\subsection{Interaction Between Energy Harvested, Storage Capacity, and Information Transfer Rate}
So far the discussion has covered the minimum energy generated by the energy harvester and software solutions that can be implemented to reduce energy consumption in order to meet that design constraint while minimally affecting performance. However, limiting the device operation to the minimal power budget will waste massive amounts of energy harvested throughout the year and potentially lose information that could have otherwise been reported. These losses can be subverted by making the device energy-aware, such that it can vary its energy consumption by altering the information transfer rate based on the amount of energy available.

The average energy consumed by the device ($E_{TxRate}$) for a specific data transmission rate can be simplified to Eq. \ref{avg_pwr_cons}. $I_{comm}$, $I_{comp}$, and $I_{off}$ is the current consumed during communication, computation, and standby mode respectively; and $T_{comm}$, $T_{comp}$, and $T_{off}$ is the time spent performing each of those tasks. As the information transfer rate or frequency of reporting samples increases, the relative value of $T_{comm}$ to the total time increases which thereby increases energy consumption.

\begin{equation}
\begin{aligned}
	E_{avg} = V(T_{comm}.I_{comm}+T_{comp}.I_{comp}+T_{off}.I_{off})
\end{aligned}
\label{avg_pwr_cons}
\end{equation}

\begin{figure}[b!]
\centering
\includegraphics[width = \columnwidth]{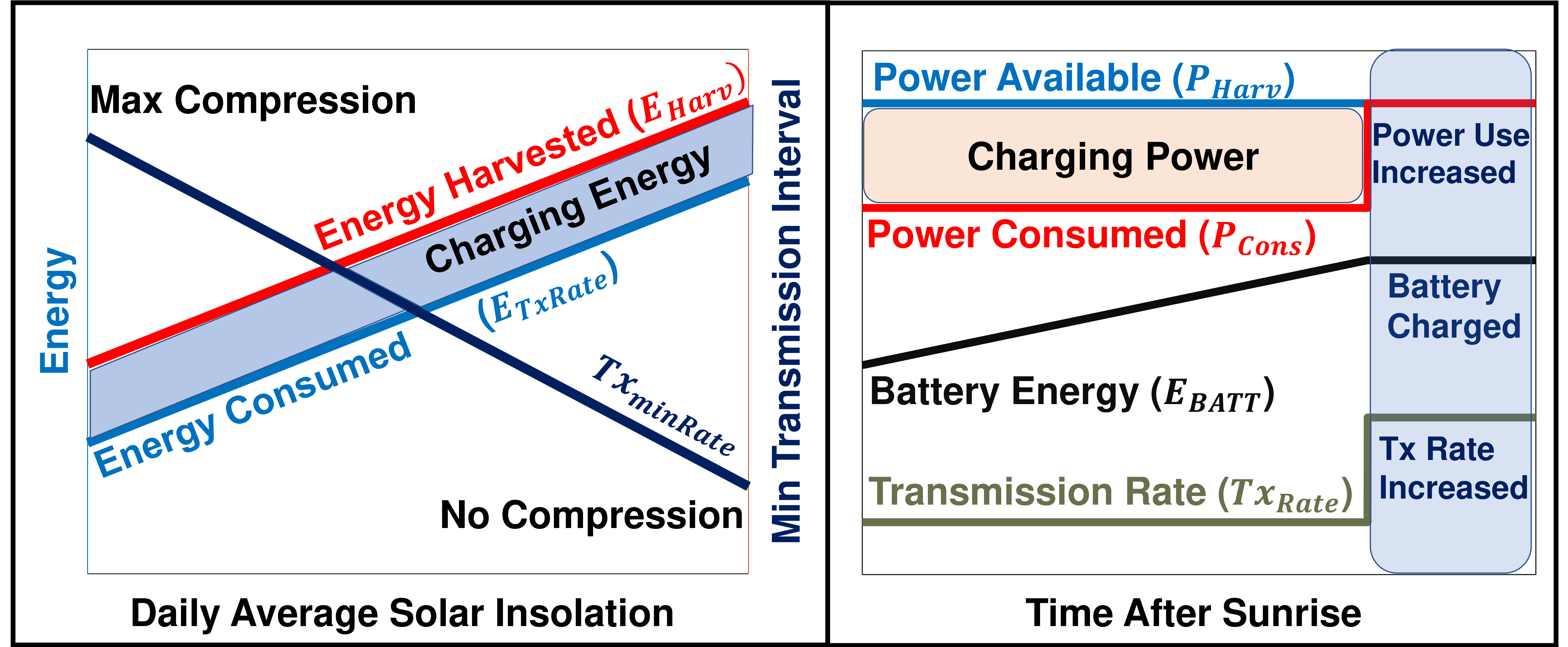}
\caption{Device behaviour shown as the relationship between the harvested energy, energy consumed by the node, charging energy or battery voltage, and long range communication transmission interval shown as a function of daily average solar insolation and time after sunrise on a particular day.}
\label{energy_tx_rel}
\end{figure}

As shown in Fig. \ref{energy_tx_rel} the energy harvested by the node increases with the increase in average solar insolation ($E_{Harv}$) received during a day. Therefore, the device can safely transmit data at a higher transmission rates throughout the entire day while keeping the total energy consumption within the bounds of the total amount of energy available. This is done by increasing the minimum transmission rate ($Tx_{MinRate}$) and the higher transmission rates will reduce the loss of information without comprising the ability of the device to perpetually function since the energy used to charge the battery will remain constant. This will translate to higher transmission rates in the summer months when more energy is available from the harvester and consequently lower transmission rates in the winter.

On a given day, during daylight hours the difference between the harvested power ($P_{Harv}$) and the power consumed by the node ($P_{Cons}$) is used to charge the energy storage device as shown in Fig. \ref{energy_tx_rel}. When the harvested power increases beyond the minimum power consumption of the device (governed by $Tx_{MinRate}$) the battery starts charging. As the sunlight intensity changes throughout the day, the data transmission rate is varied such that the amount of charging power remains constant. If the energy storage device ($E_{BATT}$) is charged to capacity, the excess power available from the energy harvester would be wasted if not consumed by the device. Therefore the power consumed by the wireless sensor node is increased to match the amount of power generated by increasing the in instantaneous data transmission rate ($Tx_{Rate}$).


\section{Platform and Implementation}

\subsection{Hardware}
\begin{figure}[t!]
\centering
\includegraphics[width = \columnwidth]{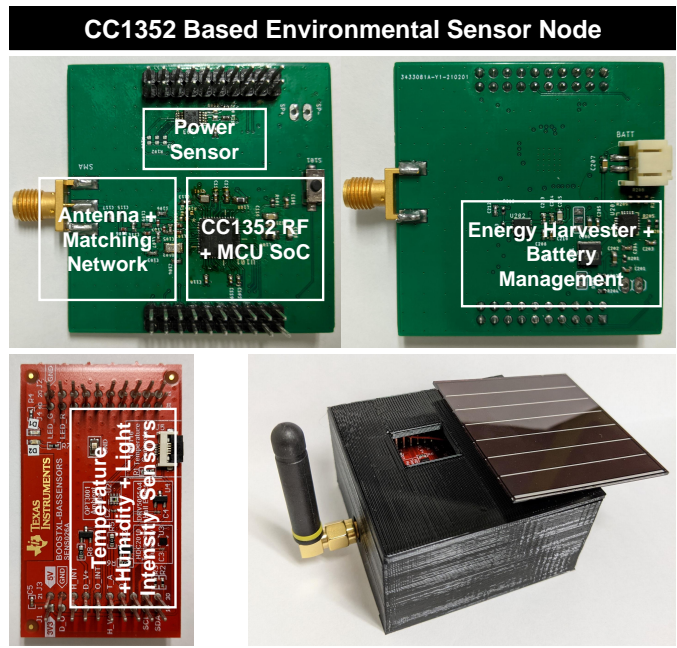}
\caption{PCB stack of the CC1352 based energy harvested long range sensor node shown along with its 3D printed housing. 50mm x 60mm amorphous silicon solar cell, with BQ25505 (TI) energy harvester used for power management. CC1352 SoC (TI) is used to implement ISA+EICO in conjunction with the power sensor (INA233). HDC2010 and OPT3001 is used as the environmental senors.}
\label{device}
\end{figure}

The custom long range sensor node shown in Fig. \ref{device} can be broadly divided into three main blocks, power management, the microcontroller and RF chain, and the environmental sensors. The device was designed to be modular and consists of two vertically stacked printed circuit boards (PCB). One PCB (top) houses only the environmental sensors to allow for easy replacement or addition of new sensors to re-purpose the device without redesigning the entire sensor node. The top layer of the bottom PCB comprises of the microcontroller and RF chain along with the power sensor and finally, the energy harvester and battery management is placed on the bottom layer.

A System-on-Chip (CC1352R1, Texas Instruments (TI)) integrates an ARM Cortex- M4F processor with a multi-band (sub-GHz and Bluetooth low energy) wireless transceiver. In this design only the sub-GHz wireless transceiver is used and the BLE transceiver is always in the powered down state. The primary motivation for selecting this SoC was minimizing power consumption while maximizing performance since it boasts of one of the lowest power architectures with high receiver sensitivity (-121 dBm for 868MHz at 5.8mA) and transmission power efficiency (+14dBm for 868MHz at 28.9mA). An integrated ultra-low power sensor controller is used to sample and process sensor data whose operation is independent of the system processor and draws 30uA at 2MHz. The system CPU consumes 2.9mA in active mode at 48MHz and 0.85uA in stand-by mode with 80KB of RAM and CPU retention, making it powerful enough to run analytics by consuming minimal power. Finally, power consumption is further reduced by using an on-chip DC-DC converter.

Three environmental variables temperature, humidity, and light intensity are collected by the sensor node over I2C. HDC2010 and OPT3001 by TI are used to measure the first two and the last quantity, respectively. HDC2010 provides data at an accuracy of 0.2 degrees Celsius for temperature and 2\% for humidity while consuming 0.55uA. OPT3001 has a measurement range of 0.01 lux to 83K lux in the visible spectrum. The sensors are powered up through a PMOS in order to turn them off during sampling intervals and conserve power.

An amorphous silicon solar cell of 60mm by 50mm is used as the power source to an ultra-low power harvester and power management IC (BQ25505, Texas Instruments). The device has cold start voltage of 600mV, consumes 325nA, performs maximum power point tracking, and can continuously harvest energy when the input voltage is as low as 100mV. The energy harvester converts the solar cell voltage to 4.2V, which is used to power the system and charge the back up battery/charge storage device. When the input power falls below the system load, an inbuilt automatic power multiplexer draws power from the charge storage device to prevent the voltage rails from drooping. The sensor node is made energy aware by measuring the battery voltage and the power drawn from the solar cell on the high side using an ultra-precise power monitor (INA233 by TI) which typically draws \( 310\ \mu \)A during normal operation and \( 2\ \mu \)A in standby mode.


\subsection{Software}

The microcontroller is programmed with an RTOS to read the sensor values every 1 second and run a network stack (Easylink by Texas Instruments) for sub-GHz communication. Data is transmitted at an interval between 1 second (no compression) to 5 minutes (maximum compression). A lightweight algorithm is implemented for anomaly detection and energy-aware data transmission to optimize power consumption and loss of information.

\begin{algorithm}
\begin{small}
\KwData{Output power of solar cell}
\KwResult{Data Transmission Rate}
initialization\;
\While{power reading available}{
    \begin{algorithmic}
    \STATE Integrate for Energy Available\;
    \IF{harvested power $>0$}{
        \IF{power reading above threshold}
        \STATE Increase data transmission rate\;
        \ELSIF{power reading below threshold}
        \STATE decrease data transmission rate up to minimum Tx rate\;
        \ELSE{
            \IF{Battery charged}
            \STATE Set data transmission rate to match available power\;
            \ENDIF}
        \ENDIF}
    \ELSE{
        \IF{Sunset Time}
            \STATE Measure battery voltage and compute energy stored\;
            \STATE Calculate new minimum transmission rate\;
            \ENDIF}
        \ENDIF
    \end{algorithmic}}
\caption{Energy-aware data transmission algorithm }
\end{small}
\end{algorithm}

The sensor node is made energy intelligent by measuring the amount of energy harvested on a given day (Eqn. \ref{en_harvest}) and the energy stored in the battery (Eqn. \ref{en_batt}) to determine the minimum energy consumption of the device for the following day. This is achieved by controlling the communication energy through adjusting the minimum transmission rate of the device, which governs the information transfer rate at nighttime and until the harvested power exceeds the power consumption of the sensor node during the day and is given by Eqn. \ref{en_avail} and Eqn. \ref{tx_min}. The minimum transmission rate is evaluated from the energy measurements once every day at sunset, when $P_{Harv}$ falls to zero for the first time. The energy harvested on a given day is shown in Eq. \ref{en_harvest}. The power sensor connected to the solar cell is sampled once every minute and numerical integration is performed to calculate the energy in Joules. Eq. \ref{en_batt} depicts the calculation of the energy stored in the battery which is performed by using a look up table on the measured battery voltage. The look-up table was generated by characterizing the battery at a discharge rate of 0.1C (23mA).

\begin{equation}
\begin{aligned}
E_{Harv} (J) &= \Sigma (P_{Harv} * 60/1000)
\end{aligned}
\label{en_harvest}
\end{equation}

\begin{equation}
\begin{aligned}
E_{BATT} (J) &= f_{Battery Chemistry} (V_{BATT})
\end{aligned}
\label{en_batt}
\end{equation}

\begin{equation}
\begin{aligned}
E_{Avail} (J) &= \frac{E_{BATT}-E_{buf}}{D_{max}} + E_{Harv}
\end{aligned}
\label{en_avail}
\end{equation}

\begin{equation}
\begin{aligned}
Tx_{minRate} &= \min_{\forall{Tx_{Rate}}}(E_{TxRate}-E_{Avail})
\end{aligned}
\label{tx_min}
\end{equation}

\begin{figure}[t!]
\centering
\includegraphics[width = \columnwidth]{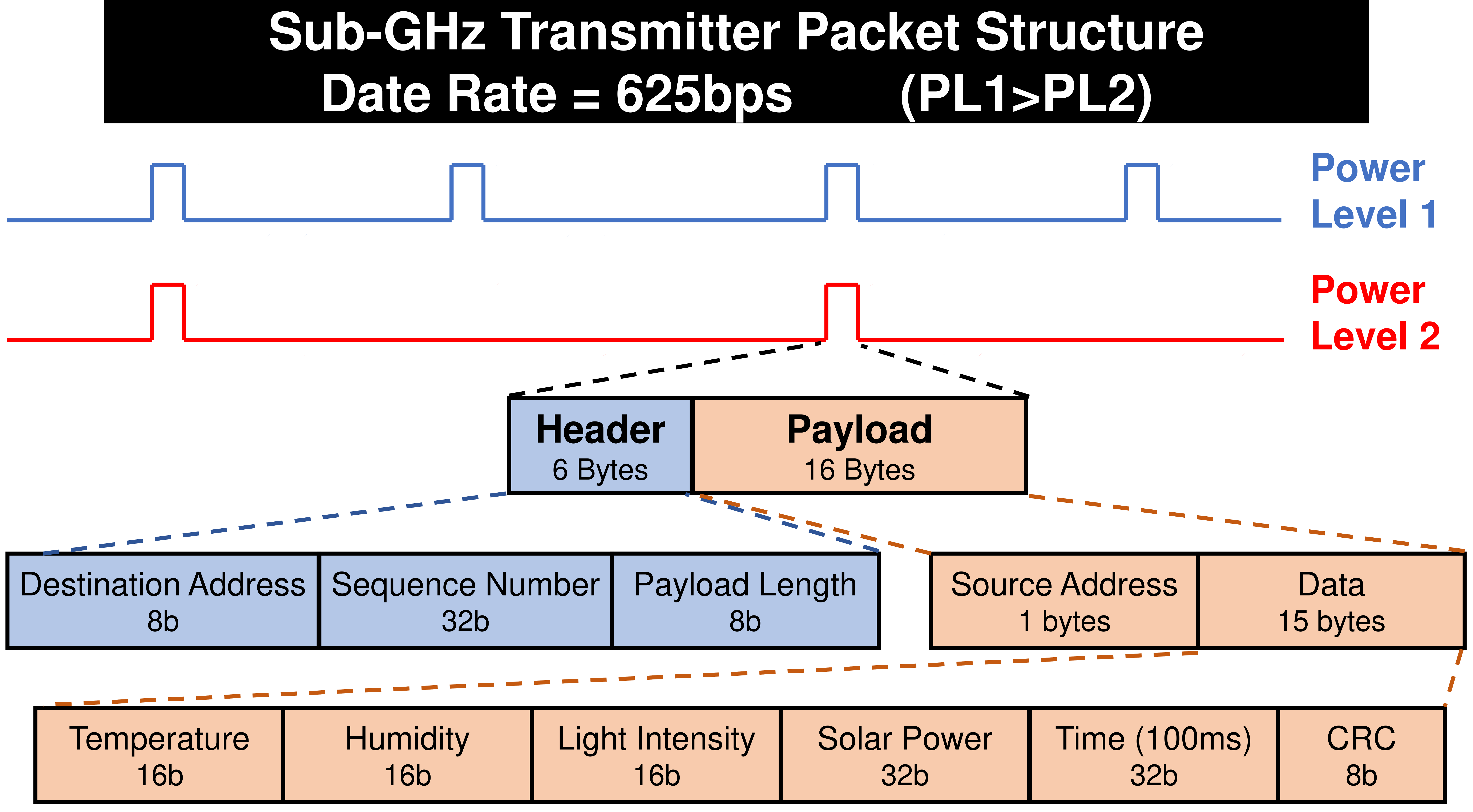}
\caption{TI Sub-GHz (Easylink) packet structure at 625 bps for long range communication. Every Tx packet 22 Bytes long with a header of 6 Bytes and payload of 16 Bytes. Packet transmission rate varies based on the total energy available to the wireless sensor node.}
\label{packet_struct}
\end{figure}

The available energy range is divided into 10 regions which maps on to a data transmission rate between 1 sample/second (no compression) to 1 sample every 300s (maximum compression). As shown in Fig. \ref{packet_struct}, each data packet is 22 bytes long carrying 16 bytes of payload. Sensor data of humidity, temperature, and light intensity is contained within 6 bytes, and 4 bytes are used to convey the time of sampling. 4 Bytes are also used to transmit information on the available power. Each device also sends its specific software defined address and error detecting codes.

\begin{algorithm}
\begin{small}
 \KwData{Samples from environmental sensors}
 \KwResult{Anomalies detected in sampled data}
 Initialize the threshold (\textit{x}) of the k-means clustering algorithm\;
 \While{New sample available}
 {
  \eIf{data $>=$ x\% or $<=$ x\% from last anomaly (ISA)}{
  activate sub-GHz communication\;
  send temporally compressed data stream to the receiver\;
  record new transmission time;
  deactivate sub-GHz communication\;
  }
  {
  wait for the next anomaly/next transmission time\;
  stay in low-power sense and compute mode\;
  }
  \eIf{time since last transmission = transmission interval}{
  activate sub-GHz communication\;
  send current data point to the receiver\;
  record new transmission time;
  deactivate sub-GHz communication\;
  }
  {
  wait for the next anomaly/next transmission time\;
  stay in low-power sense and compute mode\;
  }
 }
 \caption{Anomaly Detection followed by data transmission using Long Range communication at a particular power level}
\end{small}
\end{algorithm}

The anomaly detection algorithm is used to minimize the loss of information when lower data transmission rates are used. It incorporates a predefined threshold for each environmental variable being sensed. When the difference between the sensed data and the previous anomaly value crosses this threshold an anomaly is registered and data from all sensors is transmitted. These thresholds were calculated offline using a k-means clustering algorithm over a span of more than 4 weeks.

\begin{figure}[t!]
\centering
\includegraphics[width = \columnwidth]{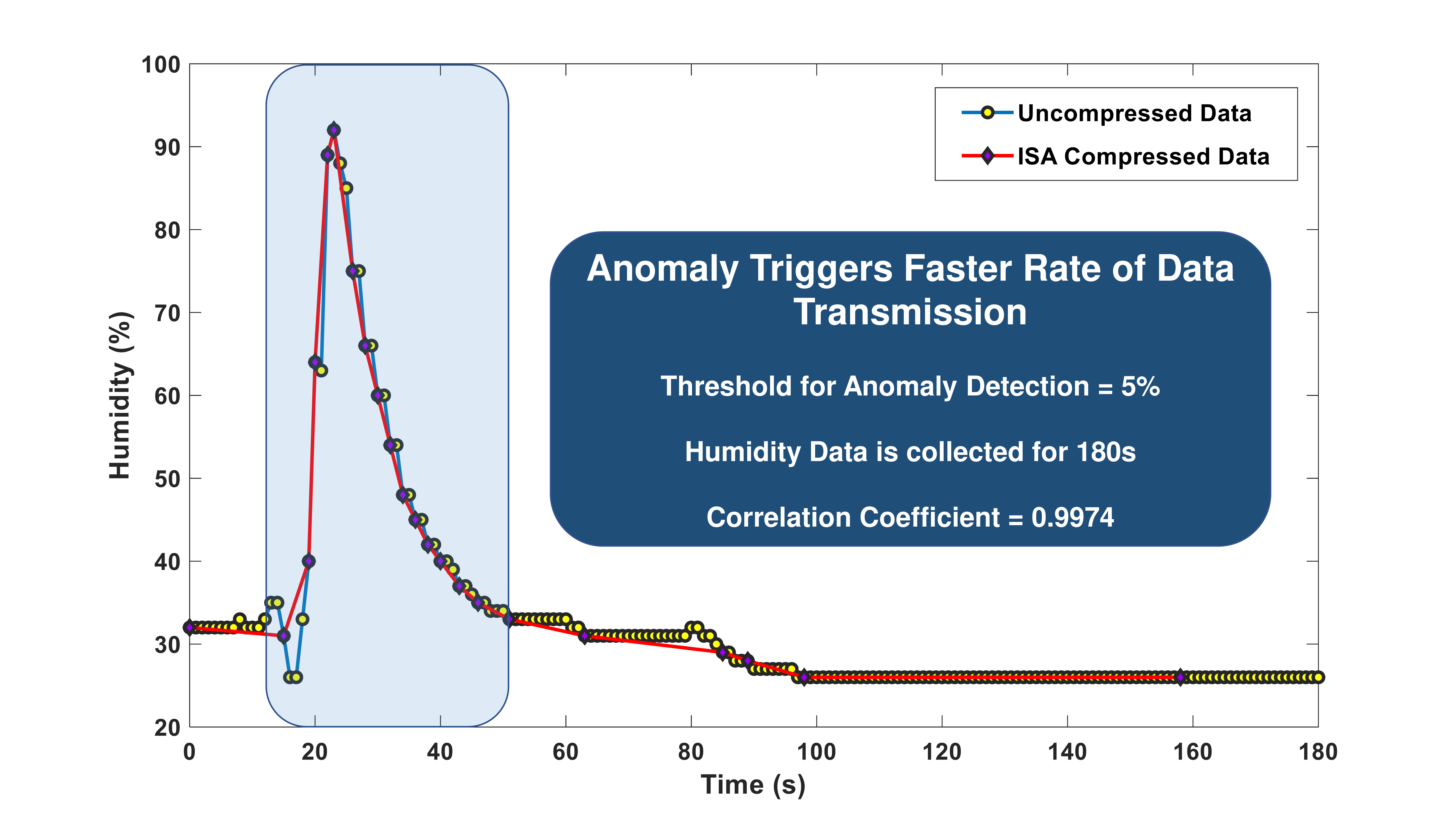}
\caption{An example of humidity data logged from HDC2010 with and without anomaly detection at a threshold of 5\%.}
\label{anomaly_example}
\end{figure}

Figure \ref{anomaly_example} shows an example of humidity data transmitted by the sensor node using anomaly detection overlaid on all of the samples collected by the microcontroller. In this case a sampling frequency of 1Hz was used and 180 data points were collected. The threshold for anomaly detection was set at 5\% with a data transmission interval of 60s. As seen a total of 23 data points were transmitted resulting in a temporal compression ratio of 7.83. Of these 23 data points, 21 were transmitted due to the anomaly created between 15s and 100s and 2 were transmitted during normal operation. Without the introduction of an anomaly the compression ratio would have been much higher. The compressed data has a correlation coefficient of 0.9974 with the original sampled data.

A report of every 1 degree Celsius change in temperature is desirable since the sensor node is primarily used for agricultural and environmental monitoring purposes. Temperatures in Indiana remain between -20 to 20 degrees Celsius for most of the year and a 5\% threshold will prevent loss of information irrespective of the data transmission interval. In the summer months when the temperatures cross 20 degrees Celsius, the anomaly detector will not trigger at every degree change in temperature. However, the sunlight intensity also increases in this time period which will result in faster minimum data transmission rates to prevent any loss of information.


\section{Results}

\subsection{Energy Consumed by the Wireless Sensor node}

\begin{figure}[t!]
\centering
\includegraphics[width = \columnwidth]{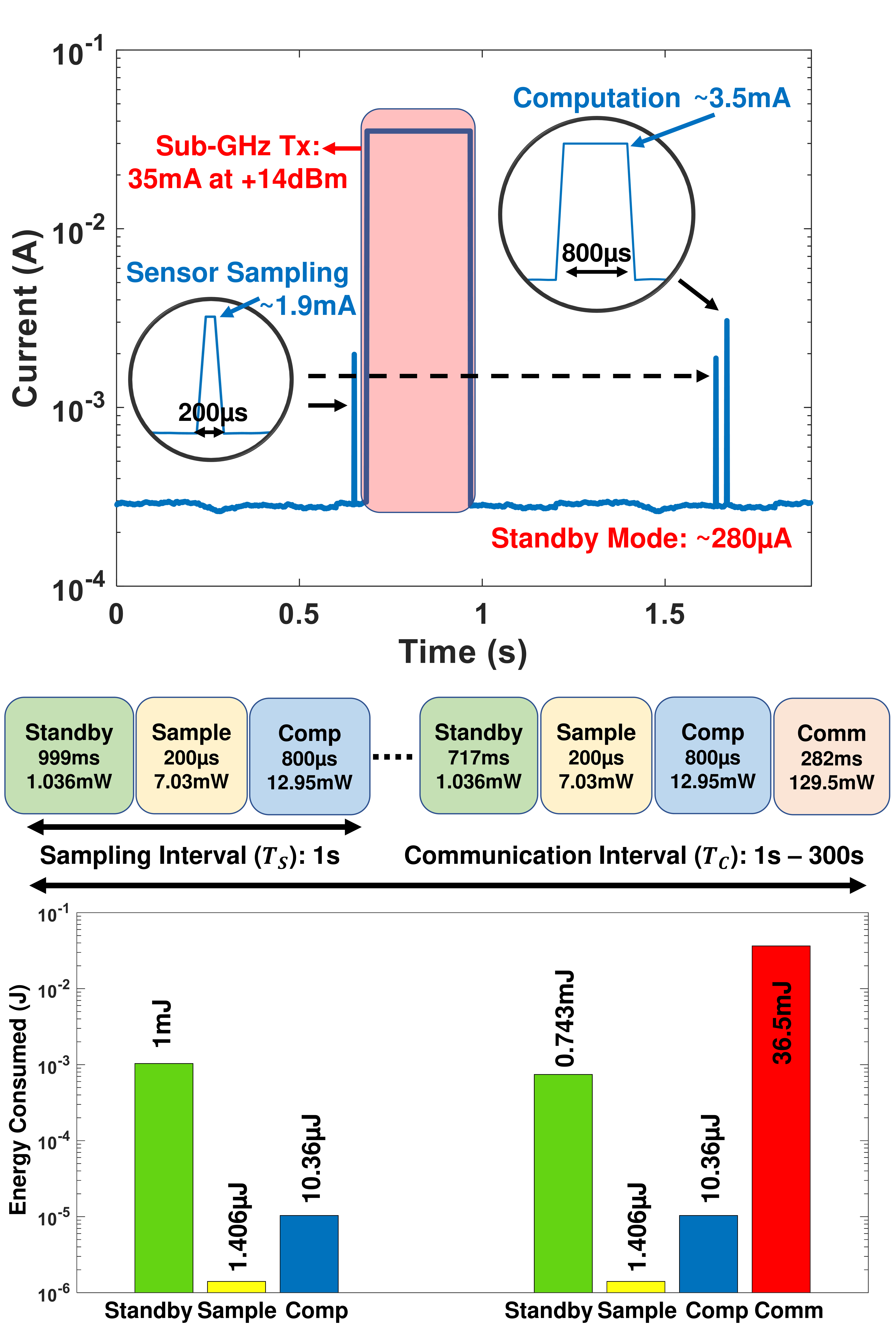}
\caption{Current consumed by the wireless sensor node as measured by a precision current-voltage analyzer and the amount of time spent and power and energy consumed in each of its different modes of operation i.e. standby (leakage), sampling and computation, and communication.}
\label{current_device}
\end{figure}

The current consumption of the sensor node in different modes of operation is measured using a precision current-voltage analyzer (B2901A, Keysight) and shown in Fig. \ref{current_device}. In standby mode the device consumes $280 \mu A$, during computation $~3.5mA$, and a peak current of $~35mA$ for sub-GHz long range transmission at an output power of +14dBm. At a supply voltage of $3.7V$, this translates to a power consumption of $1.036mW$, $7.03mW$, $12.95mW$, and $129.5mW$ during standby, sampling, computation, and communication respectively. During each sampling interval, which repeats every 1 second, the microcontroller SoC spends approximately $999ms$ in standby mode, $200 \mu s$ to sample the sensors, and $800 \mu s$ to implement the various algorithms, resulting in an energy consumption of $1.04mJ$. Each computation interval can vary between 1s to 300s, during which the SoC spends $282ms$ transmitting the sub-GHZ RF packet at the cost of standby time, resulting in a communication energy of $33.25mJ$. The standby (leakage) current is relatively high since an ultra-low noise, high PSSR, RF, low-dropout linear regulator was selected for the design which had a typical ground pin current of $~265 \mu A$. The current consumption and consequently the energy consumption can be driven down by selecting an alternate voltage regulator, however, we did not make this choice since the energy goal of our design was met.

\begin{figure}[h!]
\centering
\includegraphics[width = \columnwidth]{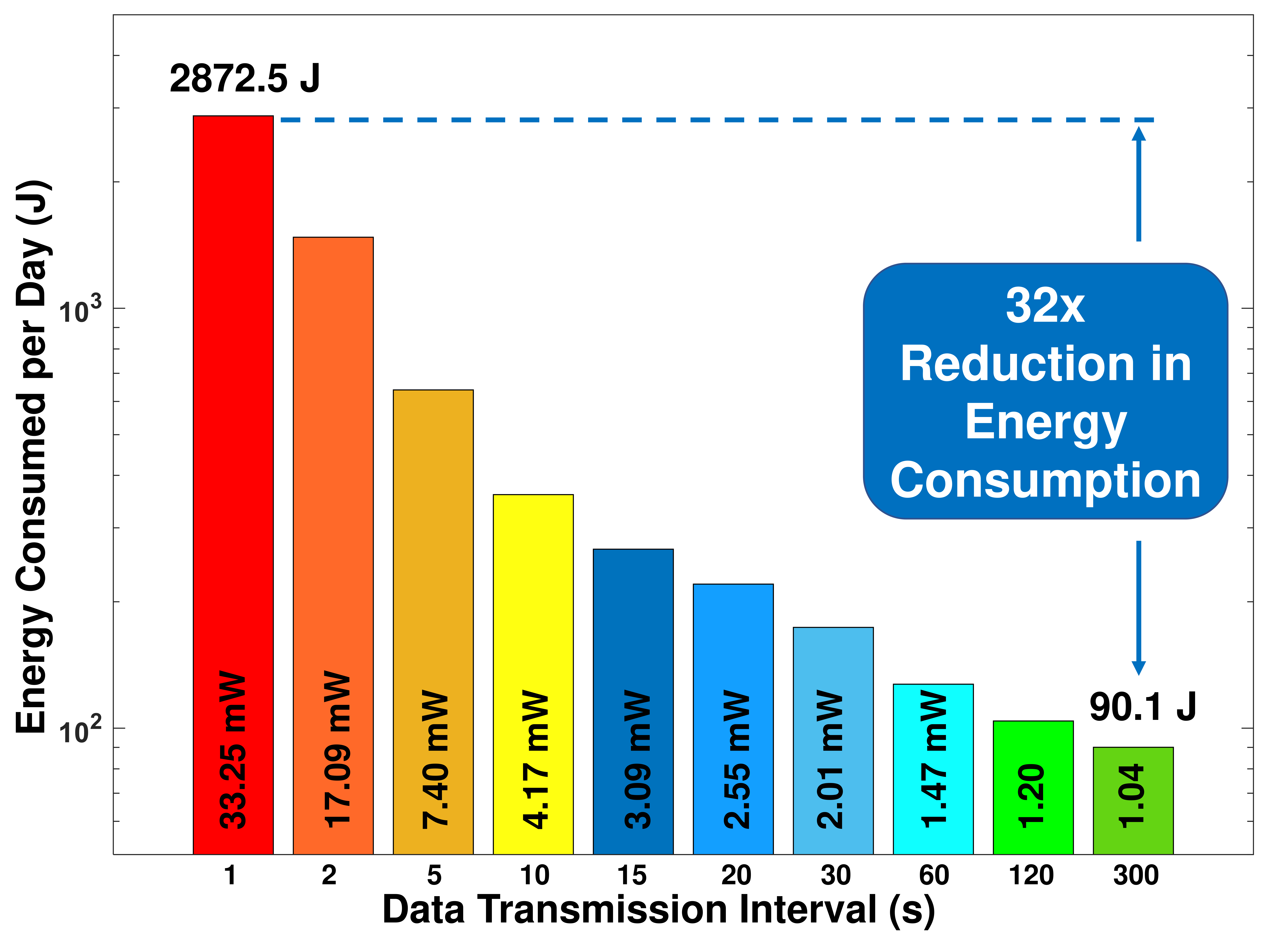}
\caption{Average energy consumed by the wireless sensor node in one day in each of its different data transmission modes used to report sensor data.}
\label{energy_profile}
\end{figure}

The energy profile for the CC1352-based energy harvested sensor node is presented in Fig. \ref{energy_profile} for each of its different data transmission intervals. The fastest data transmission rate of every 1 second (no compression) occurs either during the highest range of sunlight intensity, or when the net available energy permits a daily energy consumption of $2872.5J$. The slowest data transmission rate of once every 300 seconds (maximum compression) occurs either when negligible amounts of energy was harvested during the previous day (due to snow accumulation, etc.) or the net available energy is at a critical value to prioritize charging of the battery. This mode consumes an average energy of $90.1J$, which translates to a 32x reduction in energy consumption with less than 5\% loss of information. In case the energy harvester is incapacitated due to excessive accumulation of snow or other unforeseen circumstances, a lifetime of between 336 hours (14 days) to 818 hours can be obtained based on the amount of energy stored in a standard $230mAh$ battery.

\subsection{Interaction Between Energy Available, Energy Consumed and Transmission Rate}

\begin{figure}[t!]
\centering
\includegraphics[width = \columnwidth]{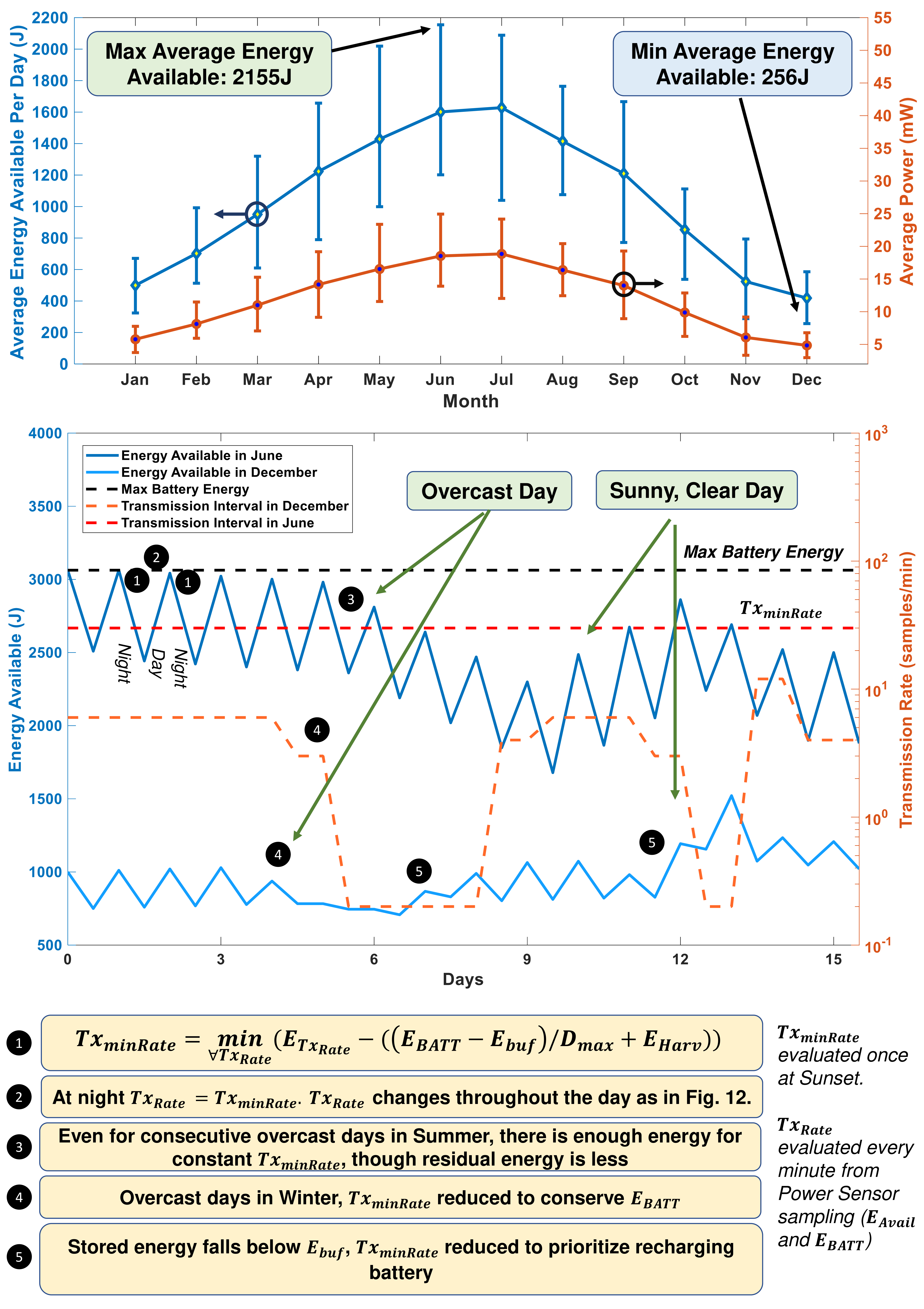}
\caption{(a) The maximum, minimum, and average power and energy available from the 50x60mm amorphous silicon solar cell in a 24-hour period on an average day of the given month in Indiana. This accounts for the losses in the energy harvester and power management system. (b) The minimum data transmission rate of the wireless sensor node as a function of energy available during a 15 day period at different times of the year.}
\label{meas_solar}
\end{figure}

The energy harvester and power management system were characterized by measuring the power available at the microcontroller supply net using a solar simulator which generated a sunlight intensity of $1KW/m^2$, also called as 1 sun or peak sun. The results obtained from this setup were multiplied by the peak sun hours seen in a day seen in Indiana during a given month to calculate the maximum, minimum, and average energy harvested on a given day of each month of the year. This is shown in Fig. \ref{meas_solar} along with the equivalent available power for a 24 hour period, such that the charge storage device sees a net zero power loss. These power measurements reflect a horizontal placement of the solar cell which will be typical during the course of using the device. Obviously, the instantaneous power available during peak sunshine hours can be up to 3X higher than the average value. The average minimum energy value of 256J in December is almost 3 times the minimum energy consumed by the device and perfectly accommodates daily fluctuations in weather and the reduction in efficiency over time due to the accumulation of dust.

\begin{equation}
\begin{aligned}
E_{Avail} (J) &= \frac{E_{BATT}-E_{buf}}{D_{max}} + E_{Harv}
\end{aligned}
\label{e_avail}
\end{equation}

\begin{equation}
\begin{aligned}
Tx_{minRate} &= \min_{\forall{Tx_{Rate}}}(E_{TxRate}-E_{Avail})
\end{aligned}
\label{en_rel}
\end{equation}

Eq. \ref{en_rel} describes the relationship between the total amount of energy available ($E_{Avail}$) which is derived from the amount of energy harvested on the previous day ($E_{Harv}$), energy stored in the battery ($E_{BATT}$), and minimum data transmission rate ($Tx_{minRate}$) which governs the energy consumed by the node. The equation assumes that at least $E_{Harv}$ will be harvested in the subsequent days and based on the amount of energy stored, it determines whether to give charging preference or allow the sensor node to burn extra energy such that it wont reach its critical threshold ($E_{Buff}$) for $D_{max}$ days. This net energy is compared to the energy consumption of each data transmission rate to find the closest match and determine the minimum data transmission rate for the following day. $E_{buf}$ represents the buffer energy in the battery (critical threshold) which must be maintained to accommodate for future bad predictions when energy availability is low.

\begin{figure}[t!]
\centering
\includegraphics[width = \columnwidth]{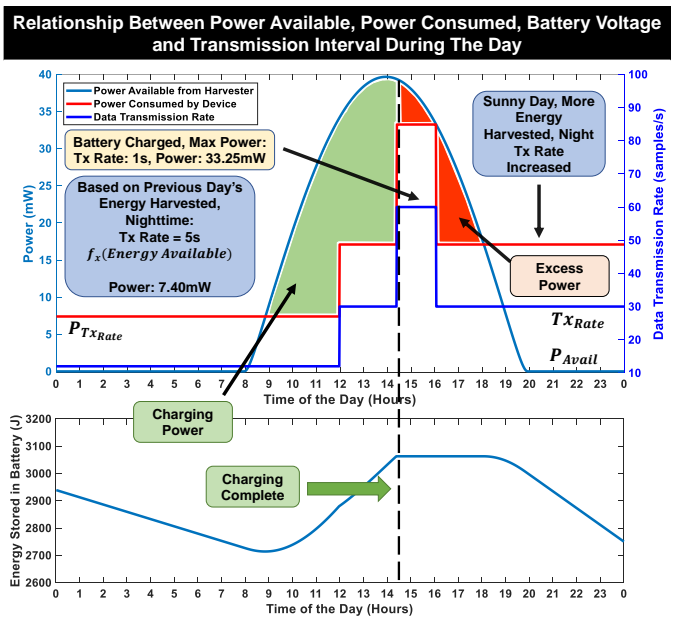}
\caption{Interaction between power available, power consumed, energy stored in the battery, and data transmission rate during the course of a sunny, clear day during March in Indiana.}
\label{power_tx_day}
\end{figure}

Fig. \ref{meas_solar} depicts the operation of the device over 15 days during both, the summer (June) and winter (December) months. The upward slopes depict charging of the battery during day time and the downward slopes for discharge during night time. During the summer months the minimum transmission rate remains steady at 30 samples/minute despite any changes in weather conditions which varies the amount of energy harvested and subsequently stored in the battery since the net energy available never moves between thresholds. An interesting point to note is that the device never enters the highest minimum transmission rate even on sunny, clear days, when maximum energy is harvested. This can be rectified by either using a larger battery, solar cell, or placing the device in a geographical location where more sunlight is available. During the winter months, the energy stored in the battery is typically lower and large fluctuations are seen in the minimum transmission rate due to changes in weather conditions to conserve the energy stored in the battery, such that the device can remain operational. When the energy stored in the battery reaches the critical threshold of $E_{buf}$, we can see the device prioritizes charging by drastically reducing the minimum transmission rate to reduce power consumption even when large amounts of energy was harvested.

During any given day, the relationship between the power available from the energy harvester, power consumed by the wireless sensor node, the energy stored, and data transmission interval is shown in Fig. \ref{power_tx_day}. This example depicts a sunny, clear day in March, in Indiana. At night time the device transmits data at the daily minimum data transmission rate which is a function of the total energy available and is depicted in Eq. \ref{en_rel}. In this example 12 samples are transmitted every minute at a power consumption of 7.40mW. Over the course of that night which was 717minutes long, 318.35J of energy was consumed. At day break, the power available from the energy harvester slowly starts to rise and eventually becomes greater than the power consumed by the device and the excess power starts to charge the battery. When the available power is 2.5X $\>$ the power consumed by the device (60\% of available power for charging), the sensor node switches to a higher data transmission rate. This continues as long as the available power increases or the battery is completely charged. Once fully charged, the device transmits data at the highest possible rate, such that its power consumption is within the bounds of available power.As the available power reduces with decreasing solar insolation, the data transmission rate decreases until it reaches the newly calculated minimum transmission rate for the following day.


\subsection{Accuracy of Data Reported at Maximum Compression}

\begin{figure}[t!]
\centering
\includegraphics[width=\columnwidth]{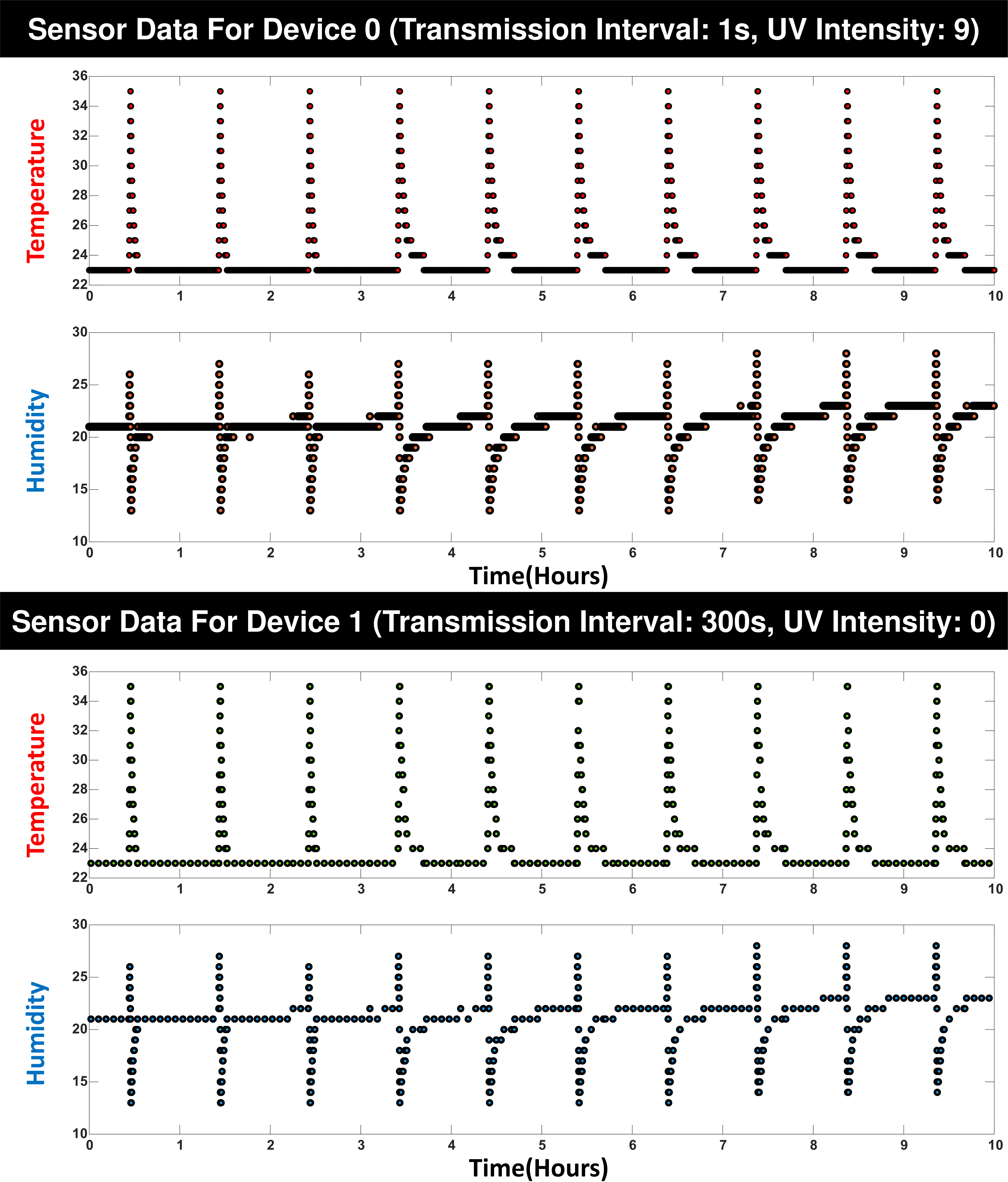}
\caption{Device 0 and Device 1 are placed at the same location with a heating pad and cooling fan placed on top of them to artificially create anomalies in their sensor readings for demonstration purposes. Device 0 operates in the maximum net energy available mode by transmitting data every second, whereas device 1 operates in the minimum net energy available mode by transmitting data every 300 seconds.}
\label{sensor_dist}
\end{figure}

Fig. \ref{sensor_dist} shows data from two sensor nodes, device 1 functioning at maximum compression and device 0 functioning at minimum compression (no compression) placed at the same location. For demonstration purposes, the devices were placed in these modes by emulating the power sensor readings to replicate the maximum and minimum power obtained from the solar cell. Additionally, to demonstrate the difference in readings reported by the devices, temperature and humidity anomalies were artificially created using a heating pad and a cooling fan which was programmed to turn on for 1 and 5 minutes respectively, once every hour. Device 1 transmits data once every 5 minutes and reports temporally compressed data when the anomaly occurs. Whereas, device 0 transmits uncompressed data every second to make maximum use of the available power. Sensor data of temperature and humidity obtained from the receiver was logged in a comma separated value (.CSV) file on a mini-PC. That data was processed and displayed in MATLAB as a time-varying quantity. A total of 418 data points were transmitted by device 1 as compared to 36000 by device 0, which resulted in a net compression ratio of 86.125 and a correlation coefficient of 0.9937 and 0.9808 for temperature and humidity respectively. This demonstration displayed a maximum energy savings of roughly 32X, due to the difference in the transmission rate of the two devices.


\section{Conclusion}
In this paper we analyzed the trade-offs and proposed the hardware design and software methods to implement a perpetually powered, energy-harvested, long-range communication sensor node $<35cm^2$ in dimension which is capable of long-range communication over $>1Km$ at continuous information transfer rates of up to 1 packet/second. This was achieved through Energy-Information Dynamic Co-optimization and in-sensor analytics. The proposed method varied the data transmission rate to optimize energy consumption based on the total amount of energy harvested and stored in the battery. This resulted in nearly continuous transmission of samples in the summer months because of large amounts of energy harvested, to a steady decrease to low data transmission rates during the winter months due to a lack of energy availability. To minimize the loss of information due to steep duty cycling of data transmission rates, ISA was employed to detect anomalies. This resulted in maximum daily energy consumption savings of approximately $32X$. Despite these transmission rate fluctuations and steep duty cycling, the correlation coefficient between the transmitted sensor data and sampled sensor data was always $>0.95$, resulting in $<5\%$ loss of information.

Although this design primarily focuses on solar power as the energy source, it can be easily modified to work with alternative sources like TEGs and piezoelectric generators to achieve an identical performance. As future work, the energy consumption would be analyzed throughout the year to ensure reliability over different weather conditions and network security can be improved. Additionally, since the leakage current of the sensor node is the limiting variable for power consumption, a custom SoC can be designed in-house to significantly reduce leakage current and further miniaturize the sensor node. Subsequently, faster data transmission rates can be achieved with smaller harvesting elements.

\ifCLASSOPTIONcaptionsoff
  \newpage
\fi



%
\bibliographystyle{ieeetr}
\small
\bibliography{bib_file}

\begin{thebibliography}{10}

\bibitem{VNI}
``{VNI Forecast Highlights Tool}.'' [Online]. Available:
  \url{https://www.cisco.com/c/m/en_us/solutions/service-provider/vni-forecast-highlights.html}.
\newblock [Accessed: Jan-30-2020].

\bibitem{habitat}
A.~Mainwaring, D.~Culler, J.~Polastre, R.~Szewczyk, and J.~Anderson, ``Wireless
  sensor networks for habitat monitoring,'' in {\em Proceedings of the 1st ACM
  International Workshop on Wireless Sensor Networks and Applications}, (New
  York, NY, USA), Association for Computing Machinery, 2002.

\bibitem{volcano}
G.~{Werner-Allen}, K.~{Lorincz}, M.~{Ruiz}, O.~{Marcillo}, J.~{Johnson},
  J.~{Lees}, and M.~{Welsh}, ``Deploying a wireless sensor network on an active
  volcano,'' {\em IEEE Internet Computing}, vol.~10, no.~2, pp.~18--25, 2006.

\bibitem{structural}
K.~Chebrolu, B.~Raman, N.~Mishra, P.~K. Valiveti, and R.~Kumar, ``Brimon: A
  sensor network system for railway bridge monitoring,'' in {\em Proceedings of
  the 6th International Conference on Mobile Systems, Applications, and
  Services}, (New York, NY, USA), Association for Computing Machinery, 2008.

\bibitem{baibhab_jiot}
B.~{Chatterjee}, D.~H. {Seo}, S.~{Chakraborty}, S.~{Avlani}, X.~{Jiang},
  H.~{Zhang}, M.~{Abdallah}, N.~{Raghunathan}, C.~{Mousoulis}, A.~{Shakouri},
  S.~{Bagchi}, D.~{Peroulis}, and S.~{Sen}, ``Context-aware collaborative
  intelligence with spatio-temporal in-sensor-analytics for efficient
  communication in a large-area iot testbed,'' {\em IEEE Internet of Things
  Journal}, pp.~1--1, 2020.

\bibitem{ehsn_survey}
S.~{Sudevalayam} and P.~{Kulkarni}, ``Energy harvesting sensor nodes: Survey
  and implications,'' {\em IEEE Communications Surveys Tutorials}, vol.~13,
  no.~3, pp.~443--461, 2011.

\bibitem{ec_eg1}
{T. Karnik et al.}, ``A cm-scale self-powered intelligent and secure iot edge
  mote featuring an ultra-low-power soc in 14nm tri-gate cmos,'' in {\em 2018
  IEEE International Solid - State Circuits Conference - (ISSCC)}, pp.~46--48,
  Feb 2018.

\bibitem{wind}
Y.~K. {Tan} and S.~K. {Panda}, ``Optimized wind energy harvesting system using
  resistance emulator and active rectifier for wireless sensor nodes,'' {\em
  IEEE Transactions on Power Electronics}, vol.~26, no.~1, pp.~38--50, 2011.

\bibitem{multi-energy}
W.~{Lee}, M.~J.~W. {Schubert}, B.~{Ooi}, and S.~J. {Ho}, ``Multi-source energy
  harvesting and storage for floating wireless sensor network nodes with long
  range communication capability,'' {\em IEEE Transactions on Industry
  Applications}, vol.~54, no.~3, pp.~2606--2615, 2018.

\bibitem{better_solar}
Z.~{Stamenkovic}, M.~{Mabon}, M.~{Gautier}, B.~{Vrigneau}, M.~{Le Gentil}, and
  O.~{Berder}, ``The smaller the better: Designing solar energy harvesting
  sensor nodes for long-range monitoring,'' {\em Wireless Communications and
  Mobile Computing}, vol.~2019, p.~11, 2019.

\bibitem{EM}
J.~{Zhang}, P.~{Li}, Y.~{Wen}, F.~{Zhang}, and C.~{Yang}, ``A management
  circuit with upconversion oscillation technology for electric-field energy
  harvesting,'' {\em IEEE Transactions on Power Electronics}, vol.~31, no.~8,
  pp.~5515--5523, 2016.

\bibitem{piezo}
N.~{Kong} and D.~S. {Ha}, ``Low-power design of a self-powered piezoelectric
  energy harvesting system with maximum power point tracking,'' {\em IEEE
  Transactions on Power Electronics}, vol.~27, no.~5, pp.~2298--2308, 2012.

\bibitem{optimal_solar}
C.~{Alippi} and C.~{Galperti}, ``An adaptive system for optimal solar energy
  harvesting in wireless sensor network nodes,'' {\em IEEE Transactions on
  Circuits and Systems I: Regular Papers}, vol.~55, no.~6, pp.~1742--1750,
  2008.

\bibitem{Ruan}
T.~{Ruan}, Z.~J. {Chew}, and M.~{Zhu}, ``Energy-aware approaches for energy
  harvesting powered wireless sensor nodes,'' {\em IEEE Sensors Journal},
  vol.~17, no.~7, pp.~2165--2173, 2017.

\bibitem{solar_prediction}
J.~{Recas Piorno}, C.~{Bergonzini}, D.~{Atienza}, and T.~{Simunic Rosing},
  ``Prediction and management in energy harvested wireless sensor nodes,'' in
  {\em 2009 1st International Conference on Wireless Communication, Vehicular
  Technology, Information Theory and Aerospace Electronic Systems Technology},
  pp.~6--10, 2009.

\bibitem{TISpec}
``{CC1352R Product Specs}.'' [Online]. Available:
  \url{https://www.ti.com/lit/ds/symlink/cc1352r.pdf?ts=1616415713151}.
\newblock [Accessed: Sep-14-2020].

\bibitem{thermo_spec}
N.~Jaziri, A.~Boughamoura, J.~Müller, B.~Mezghani, F.~Tounsi, and M.~Ismail,
  ``A comprehensive review of thermoelectric generators: Technologies and
  common applications,'' {\em Energy Reports}, vol.~6, pp.~264--287, 2020.

\bibitem{rf_spec}
V.~Kuhn, C.~Lahuec, F.~Seguin, and C.~Person, ``A multi-band stacked rf energy
  harvester with rf-to-dc efficiency up to 84\%,'' {\em IEEE Transactions on
  Microwave Theory and Techniques}, vol.~63, no.~5, pp.~1768--1778, 2015.

\bibitem{solar_spec}
M.~A. Green, K.~Emery, Y.~Hishikawa, W.~Warta, E.~D. Dunlop, D.~H. Levi, and
  A.~W.~Y. Ho-Baillie, ``Solar cell efficiency tables (version 49),'' {\em
  Progress in Photovoltaics}, vol.~25, 11 2016.

\bibitem{nasa}
A.~H. Sparks, ``nasapower: A nasa power global meteorology, surface solar
  energy and climatology data client for r,'' {\em The Journal of Open Source
  Software}, vol.~3, p.~1035, oct 2018.

\bibitem{NTV}
B.~{Chatterjee}, P.~{Panda}, S.~{Maity}, A.~{Biswas}, K.~{Roy}, and S.~{Sen},
  ``Exploiting inherent error resiliency of deep neural networks to achieve
  extreme energy efficiency through mixed-signal neurons,'' {\em IEEE
  Transactions on Very Large Scale Integration (VLSI) Systems}, vol.~27,
  pp.~1365--1377, June 2019.

\bibitem{lndr}
C.~H. Bennett, ``Notes on landauer's principle, reversible computation, and
  maxwell's demon,'' {\em Studies in History and Philosophy of Science Part B:
  Studies in History and Philosophy of Modern Physics}, vol.~34, no.~3, pp.~501
  -- 510, 2003.
\newblock Quantum Information and Computation.

\bibitem{Friis_eqn}
H.~T. Friis, ``{A Note on a Simple Transmission Formula},'' {\em Proceedings of
  the IRE}, vol.~34, pp.~254--256, May 1946.

\bibitem{Friis_link_analysis}
A.~J. Johansson, ``{Performance of a radio link between a base station and a
  medical implant utilising the MICS standard},'' in {\em The 26th Annual
  International Conference of the IEEE Engineering in Medicine and Biology
  Society}, vol.~1, pp.~2113--2116, Sep. 2004.

\bibitem{CompvsComm}
B.~{Chatterjee}, N.~{Cao}, A.~{Raychowdhury}, and S.~{Sen}, ``Context-aware
  intelligence in resource-constrained iot nodes: Opportunities and
  challenges,'' {\em IEEE Design Test}, vol.~36, pp.~7--40, April 2019.

\bibitem{30pJb}
A.~Ebrazeh and P.~Mohseni, ``{30 pJ/b, 67 Mbps, Centimeter-to-Meter Range Data
  Telemetry With an IR-UWB Wireless Link},'' {\em IEEE Transactions on
  Biomedical Circuits and Systems}, vol.~9, pp.~362--369, June 2015.

\bibitem{ec_eg2}
{S. Paul et al.}, ``An energy harvesting wireless sensor node for iot systems
  featuring a near-threshold voltage ia-32 microcontroller in 14nm tri-gate
  cmos,'' in {\em 2016 IEEE Symposium on VLSI Circuits (VLSI-Circuits)},
  pp.~1--2, June 2016.

\end{thebibliography}

%

\begin{IEEEbiography}[{\includegraphics[width=1in,height=1.25in,clip,keepaspectratio]{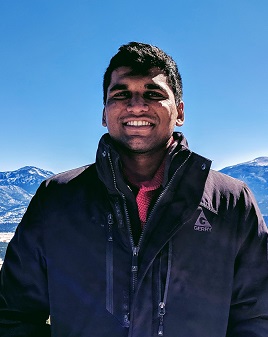}}]
{Shitij Avlani}
received the Bachelor of Engineering in Electronics Engineering from the University of Mumbai, India in 2017, during which time he worked at IIT-Bombay for three years on point of care medical devices and fabrication of MEMS sensors. He is currently working towards a M.S degree in the School of Electrical Engineering at Purdue University, West Lafayette, IN, USA. His research interests include embedded systems and analog and mixed signal circuits for sensor nodes.
\end{IEEEbiography}

\begin{IEEEbiography}[{\includegraphics[width=1in,height=1.25in,clip,keepaspectratio]{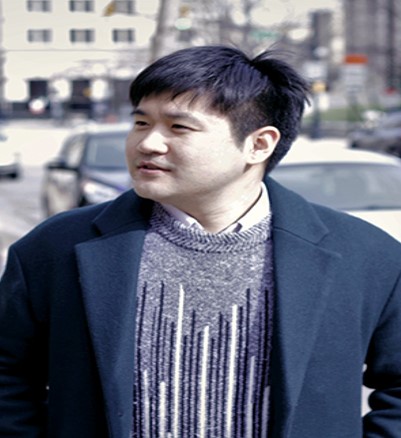}}]
{Donghyun Seo}
received the B.S. degree in electronics and radio engineering from Kyung Hee University, Seoul, South Korea, in 2013, and the M.S. degree in electronics computer engineering at Hanyang University, Seoul, South Korea, in 2015, and is currently working toward Ph.D. degree in the School of Electrical Engineering, Purdue University, West Lafayette, IN, USA. His research interests include CMOS low-power analog, mixed signal and RF integrated circuit design for sensor node interfacing.
\end{IEEEbiography}

\begin{IEEEbiography}[{\includegraphics[width=1in,height=1.25in,clip,keepaspectratio]{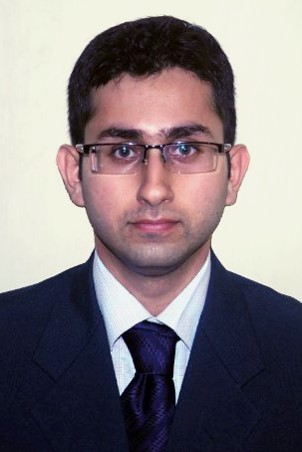}}]
{Baibhab Chatterjee}
 (S`17) received the B.Tech. degree in electronics and communication engineering from the National Institute of Technology (NIT), Durgapur, India, in 2011, and the M.Tech. degree in electrical engineering from IIT Bombay, Mumbai, India, in 2015. He is currently pursuing the Ph.D. degree with the School of Electrical Engineering, Purdue University, West Lafayette, IN, USA. His industry experience includes two years as a Digital Design Engineer/a Senior Digital Design Engineer with Intel, Bengaluru, India, and one year as a Research and Development Engineer with Tejas Networks, Bengaluru. His research interests include low-power analog, RF, and mixed-signal circuit design for secure biomedical applications.
 
 Mr. Chatterjee received the University Gold Medal from NIT, Durgapur, India, in 2011, the Institute Silver Medal from IIT Bombay in 2015, the Andrews Fellowship at Purdue University during 2017-2019, the HOST 2018 Best Student Poster Award (3rd), the CICC 2019 Best Paper Award (overall) and the RFIC/IMS 2020 3MT Award (audience choice).\\
\end{IEEEbiography}

\vspace*{\fill}
\begin{IEEEbiography}[{\includegraphics[width=1in,height=1.25in,clip,keepaspectratio]{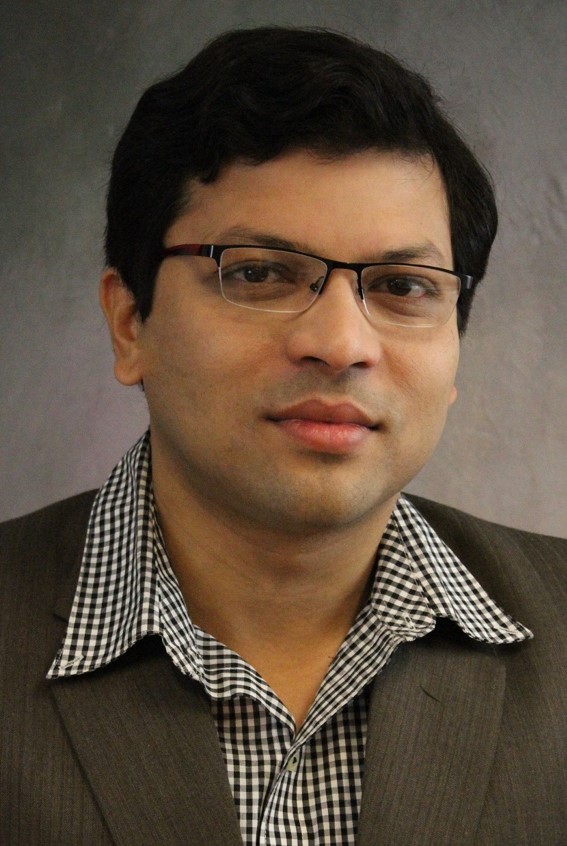}}]
{Shreyas Sen}
(S`06-M`11-SM`17) is an Associate Professor in ECE, Purdue University and received his Ph.D. degree in ECE, Georgia Tech. Dr. Sen has over 5 years of industry research experience in Intel Labs, Qualcomm and Rambus. His current research interests span mixed-signal circuits/systems and electromagnetics for the Internet of Things (IoT), Biomedical, and Security. Dr. Sen is the inventor of the Electro-Quasistatic Human Body Communication, for which, he is the recipient of the MIT Technology Review top-10 Indian Inventor Worldwide under 35 (MIT TR35 India) Award. Dr. Sen's work has been covered by 100+ news releases worldwide, invited appearance on TEDx Indianapolis, Indian National Television CNBC TV18 Young Turks Program and NPR subsidiary Lakeshore Public Radio. Dr. Sen is a recipient of the NSF CAREER Award 2020, AFOSR Young Investigator Award 2016, NSF CISE CRII Award 2017, Google Faculty Research Award 2017, Intel Labs Quality Award for industrywide impact on USB-C type, Intel Ph.D. Fellowship 2010, IEEE Microwave Fellowship 2008 and seven best paper awards including IEEE CICC 2019 and IEEE HOST  2017, 2018, and 2019. Dr. Sen's work was chosen as one of the top-10 papers in the Hardware Security field over the past 6 years (TopPicks 2019). He has co-authored 2 book chapters, over 135 journal and conference papers, and has 14 patents granted/pending. He serves/has served as an Associate Editor for IEEE Design \& Test, Executive Committee member of IEEE Central Indiana Section and Technical Program Committee member of DAC, CICC, DATE, ISLPED, ICCAD, ITC, VLSI Design, among others. Dr. Sen is a Senior Member of IEEE.
\end{IEEEbiography}




\end{document}